\documentclass[twocolumn]{aastex631}

\usepackage{array}
\usepackage{amsmath,bm}
\usepackage{longtable}
\usepackage{natbib}
\usepackage{xcolor}

\usepackage[T1]{fontenc}
\usepackage{newtxtext,newtxmath}

\DeclareRobustCommand{\VAN}[3]{#2}
\let\VANthebibliography\thebibliography
\def\thebibliography{\DeclareRobustCommand{\VAN}[3]{##3}\VANthebibliography}

\usepackage{graphicx}	
\usepackage{bm}
\usepackage{amsmath}	
\usepackage{booktabs}
\usepackage{gensymb}

\begin{document}

\renewcommand{\arraystretch}{1.5} 

\title[NGC 5005]{The X-ray Emission of NGC~5005: An Unobscured Low-Luminosity AGN with a Weakly Accreting Broad-Line Region}

\correspondingauthor{Anna Trindade Falcão}
\email{annatrindadefalcao@gmail.com}

\author{Anna Trindade Falcão} 
\affiliation{NASA-Goddard Space Flight Center, Code 662, Greenbelt, MD 20771, USA}
\affiliation{Harvard-Smithsonian Center for Astrophysics, 60 Garden St., Cambridge, MA 02138, USA}

\author{R. Middei}
\affiliation{Harvard-Smithsonian Center for Astrophysics, 60 Garden St., Cambridge, MA 02138, USA}
\affiliation{INAF Osservatorio Astronomico di Roma, Via Frascati 33, 00078 Monte Porzio Catone, RM, Italy}
\affiliation{Space Science Data Center, Agenzia Spaziale Italiana, Via del Politecnico snc, 00133 Roma, Italy}

\author{G. Fabbiano}
\affiliation{Harvard-Smithsonian Center for Astrophysics, 60 Garden St., Cambridge, MA 02138, USA}

\author{M. Elvis}
\affiliation{Harvard-Smithsonian Center for Astrophysics, 60 Garden St., Cambridge, MA 02138, USA}

\author{P. Zhu}
\affiliation{Harvard-Smithsonian Center for Astrophysics, 60 Garden St., Cambridge, MA 02138, USA}

\author{W. P. Maksym}
\affiliation{NASA Marshall Space Flight Center, Huntsville, AL 35812, USA}

\author{D.~{\L}.~Kr\'{o}l}
\affiliation{Harvard-Smithsonian Center for Astrophysics, 60 Garden St., Cambridge, MA 02138, USA}

\author{L. Feuillet}
\affiliation{Institute for Astrophysics and Computational Sciences \\
and Department of Physics, The Catholic University of America, \\
Washington, DC 20064, USA}

\begin{abstract}
We present deep \textit{Chandra} X-ray observations of NGC~5005, a LINER-dominated galaxy previously reported to host a broad H$\alpha$ emission line. The diffuse soft X-ray emission ($<$3~keV) extends out to $\sim$800~pc, while harder emission ($>$3~keV) is confined to the central $\sim$400~pc. Spatially resolved spectroscopy of the nuclear ($r<150$~pc) and extended ($150<r<500$~pc) regions reveals that these are best described by models including both photoionized and thermal plasma components, consistent with excitation by a low-luminosity AGN and shock-heated gas. Narrow-band imaging and excitation maps from the \textit{Hubble Space Telescope (HST)} support this interpretation, closely matching the X-ray morphology and ionization structure. The detection of a faint hard X-ray nuclear source with \textit{Chandra}, combined with stringent upper limits from \textit{NuSTAR} and \textit{Swift}, and consistency with the X-ray luminosity predicted from the \textit{HST} [O~III]$\lambda$5007 emission, indicates that NGC~5005 hosts an intrinsically low-luminosity ($L_{\rm bol} \sim 10^{41}$~erg~s$^{-1}$), unobscured AGN. Despite the extremely low Eddington ratio inferred from our measurements ($\lambda_{\rm Edd} \sim 5 \times 10^{-6}$), the presence of a broad H$\alpha$ line in the optical spectrum suggests the persistence of a thin accretion disk, challenging standard paradigms of accretion flow configurations at such low accretion rates.
\end{abstract}



\section{Introduction}
\label{sec:introduction}

This paper is the second in our series on the LINER active galactic nucleus (AGN) galaxy NGC~5005, and the first to present results based on deep \textit{Chandra} observations of its nuclear region. NGC~5005 is a weakly barred spiral galaxy [SAB(r)] with a dust-enshrouded nucleus \citep{pogge_narrow-line_2000}, located at a distance of $D\sim$17~Mpc (NED; $z$=0.003156, scale = 100~pc~arcsec$^{-1}$).

Early optical observations revealed strong nuclear emission lines, including [O~II], [O~III], H$\alpha$, and [S~II] \citep{shuder_empirical_1981}, but did not establish a definitive AGN classification. While initially considered unlikely to be a Seyfert 2, subsequent studies have variously classified NGC~5005 as a low-ionization nuclear emission-line region (LINER; \citealt{huchra_spatial_1992}; \citealt{ho_search_1997}; \citealt{veron-cetty_catalogue_2006}), a Seyfert 2 galaxy \citep{audibert_probing_2017}, or a hybrid of both types \citep{spinelli_atlas_2006, malkan_emission_2017}.

Spectroscopy from the Palomar survey identified a broad H$\alpha$ component blended with narrow H$\alpha$ and [N~II] lines, suggesting the presence of an unobscured AGN \citep{rush_soft_1996, ho_search_1997}. However, follow-up \textit{Hubble Space Telescope (HST)} spectroscopy using the [O~I] line for deblending did not confirm the broad component. This led \citet{balmaverde_hst_2014} to propose either a misidentification or a possible changing-look AGN. In contrast, \citet{constantin_dissecting_2015} detected a broad H$\alpha$ component (FWHM = 2,610~km~s$^{-1}$) using [S~II] as the deblending template. \citet{cazzoli_optical_2018} later confirmed the presence of a weak broad H$\alpha$ line (FWHM = 2,152~km~s$^{-1}$) in the \textit{HST} data, although it remained undetected in ground-based spectra.

The presence of a Compton-thick (CT) AGN was initially proposed by \citet{risaliti_distribution_1999} based on a low 2–10~keV to [O~III] luminosity ratio. However, their \textit{ASCA} spectrum lacked a strong 6.4~keV Fe~K$\alpha$ line, placing an upper limit of 900~eV on its equivalent width (EW). This led to the interpretation that soft X-rays were fully obscured and the observed spectrum was dominated by star-formation-related emission.

Subsequent observations with \textit{Chandra} and \textit{XMM-Newton} detected extended soft X-ray emission aligned with the galaxy disk \citep{guainazzi_transmission-dominated_2005}, likely contributing to the excess emission observed around 0.6–1~keV \citep{gallo_xmmnewton_2006}. \citet{guainazzi_transmission-dominated_2005} ruled out an inverse Compton origin for the soft component, placing a tighter upper limit of 240~eV on the Fe~K$\alpha$ EW, and favored a significantly lower absorbing column ($N_{\mathrm{H}} \sim 1.5 \times 10^{20}$~cm$^{-2}$), revising the CT interpretation. Later studies reported $N_{\mathrm{H}}$ values in the range of $10^{20}$–$10^{21}$~cm$^{-2}$ \citep{brightman_xmmnewton_2011, younes_study_2011}.

In \citet{trindade_falcao_mapping_2025}, we analyzed continuum-subtracted \textit{HST} narrow-line images to explore the spatially resolved excitation structure in the narrow-line region (NLR) of NGC~5005. We identified a compact ($\sim$100~pc) nuclear Seyfert-like zone, likely photoionized by the AGN, surrounded by a higher ionization LINER-like cocoon indicative of shock excitation in the interstellar medium (ISM). Additional LINER-like emission extends across the central kiloparsec, consistent with ionization from a low-luminosity AGN. We also detected H~II (star-forming)-like regions within the inner 500~pc, possibly arising from jet–ISM interactions, and in a star-forming ring at $r \sim 4$~kpc.

In this paper, we examine the nuclear and extended X-ray emission of NGC~5005 using the full 255~ks of archival \textit{Chandra} ACIS-S observations, in combination with archival \textit{NuSTAR} and \textit{Swift} data. We focus on the spatial and spectral properties of the X-ray emission to assess the AGN’s luminosity, absorption properties, and its role in exciting the surrounding ISM. Section~\ref{sec:Observations, Data Reduction, Analysis} describes the observations, data reduction, and analysis procedures. In Sections~\ref{sec:imaging_analysis_results} and \ref{sec:spectral_analysis_results}, we present results from imaging and spectral analysis, respectively. Section~\ref{sec:discussion} discusses the implications of our findings, and Section~\ref{sec:conclusions} summarizes our conclusions.

NGC~5005 thus offers a uniquely valuable case for probing the duty cycle and feedback signatures of LINER-type AGNs. Its comparatively faint X-ray nucleus ($L_{\rm X}\sim10^{40}$~erg~s$^{-1}$; \citealt{gonzalez-martin_x-ray_2009}) contrasts with its luminous [O~III] emission \citep{trindade_falcao_mapping_2025}, raising long-standing questions about the true nature of the central source. This ambiguity, combined with its proximity ($D \sim 17$~Mpc), allows \textit{Chandra} to resolve $\sim$30~pc scales and directly connect nuclear activity to circumnuclear excitation. Equally important, NGC~5005 is exceptionally well studied across the spectrum, with available \textit{HST} narrow-band imaging, radio jet detections \citep{baldi_lemmings_2018}, circumnuclear CO structures \citep{sakamoto_gasdynamics_2000}, and rich optical/IR data on dust, gas, and star formation \citep[e.g.,][]{pogge_narrow-line_2000, asmus_subarcsecond_2014}. These factors make it a natural choice for a detailed case study, providing both technical advantages and a pioneering framework for broader population studies of LINER-dominated systems.

\section{Observations, Data Reduction, and Analysis}
\label{sec:Observations, Data Reduction, Analysis}
\subsection{Chandra Observations}
\label{sec:observations_analysis}

Table~\ref{tab:xray_observations} summarizes the X-ray observations analyzed in this study. NGC~5005 was observed a total of sixteen times with \textit{Chandra}, primarily during the 2022-2023 observing campaign, resulting in a cumulative exposure of $\sim$255~ks. All \textit{Chandra} datasets used in this work are publicly available through the \textit{Chandra} Data Collection (CDC) 358~\dataset[doi:10.25574/cdc.358]{https://doi.org/10.25574/cdc.358}.

Data reduction was carried using \texttt{CIAO} 4.17\footnote{\url{https://cxc.cfa.harvard.edu/ciao/}} \citep{fruscione_ciao_2006}, following standard procedures. Each observation was reprocessed with the \texttt{chandra\_repro}\footnote{\url{https://cxc.cfa.harvard.edu/ciao/ahelp/chandra\_repro.html}} script, which applies the latest calibrations and enables sub-pixel resolution capabilities specific to ACIS-S.

For imaging analysis, individual datasets were aligned and merged using ObsID~25698, the longest single exposure, as the astrometric reference. Full-band (0.3–7~keV) images from each observation were visually inspected to check alignment quality. Centroid shifts between datasets were measured using the \texttt{CIAO} tool \texttt{dmstat}\footnote{\url{https://cxc.cfa.harvard.edu/ciao/ahelp/dmstat.html}}, applied to 0.3–7~keV images binned at 1/8 the native pixel size. Centroids were calculated within a $r=0.2''$ circular region centered on the brightest pixel. All observed centroid shifts were smaller than the native ACIS-S pixel scale of $0.492''$.
 
\begin{figure*}
  \centering
  \includegraphics[width=\textwidth]{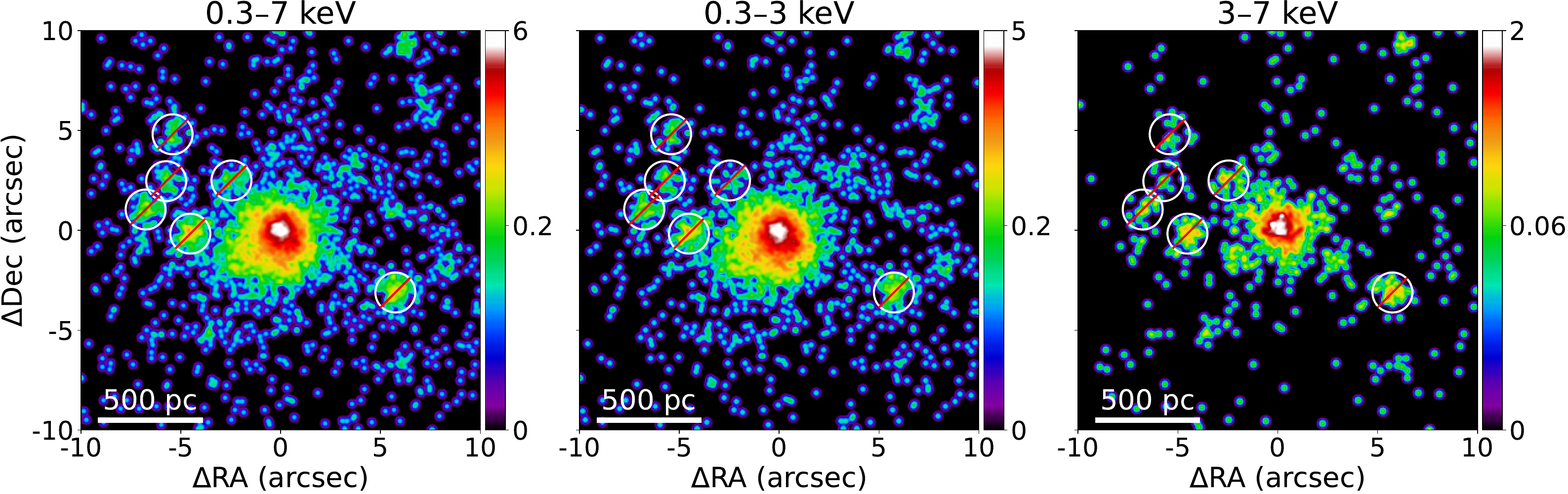}

\caption{Merged 1/8 subpixel \textit{Chandra}/ACIS-S images of NGC~5005, smoothed with a 3-pixel Gaussian kernel and shown in the indicated energy bands. Non-nuclear point sources, which were excluded from the spatial and spectral analysis, are marked. In this and all other images displayed in this paper, north is up and east is to the left, with color bars shown in units of counts.}
 \label{fig:merged_image} 
 \end{figure*}

To ensure astrometric consistency across exposures, we cross-matched additional point sources and corrected any residual offsets using \texttt{wcs\_update}\footnote{\url{https://cxc.cfa.harvard.edu/ciao/ahelp/wcs\_update.html}}. Updated aspect solutions were incorporated using \texttt{reproject\_obs}\footnote{\url{https://cxc.cfa.harvard.edu/ciao/ahelp/reproject\_obs.html}}, and final merged datasets were generated with \texttt{merge\_obs}\footnote{\url{https://cxc.cfa.harvard.edu/ciao/ahelp/merge\_obs.html}}. Merged images in the full (0.3–7~keV), soft (0.3–3~keV), and hard (3–7~keV) energy bands are shown in Figure~\ref{fig:merged_image}. These images were binned at 1/8 of the native pixel scale and smoothed with a 3-pixel Gaussian kernel to take advantage of sub-pixel event repositioning, enhancing spatial detail for high-resolution morphological analysis while suppressing pixel-scale noise fluctuations.

\begin{table}
\centering
\caption{X-ray Observations of NGC~5005 used in this work.}
\label{tab:xray_observations}
\begin{tabular}{cccc}
\toprule
\textbf{Obs ID} & \textbf{Date} & \textbf{Exposure (ks)} & \textbf{PI} \\
\midrule
\multicolumn{4}{c}{\textbf{\textit{Chandra} ACIS-S}} \\
\midrule
4021   & 2003-08-19 & 5.0   & Satyapal \\
25255  & 2023-07-19 & 15.0  & Fabbiano \\
25694  & 2023-08-06 & 14.0  & Fabbiano \\
25695  & 2022-08-09 & 14.0  & Fabbiano \\
25696  & 2023-03-04 & 12.0  & Fabbiano \\
25697  & 2023-08-06 & 10.0  & Fabbiano \\
25698  & 2023-06-16 & 44.0  & Fabbiano \\
25699  & 2023-06-18 & 16.0  & Fabbiano \\
25700  & 2023-05-12 & 30.0  & Fabbiano \\
27251  & 2022-08-11 & 14.0  & Fabbiano \\
27252  & 2023-08-12 & 10.5  & Fabbiano \\
27958  & 2023-07-19 & 14.0  & Fabbiano \\
27976  & 2023-08-22 & 16.0  & Fabbiano \\
28372  & 2023-08-12 & 12.0  & Fabbiano \\
28493  & 2023-08-22 & 18.5  & Fabbiano \\
28494  & 2023-08-23 & 10.0  & Fabbiano \\
\midrule
\multicolumn{4}{c}{\textbf{\textit{NuSTAR} FPMA/FPMB}} \\
\midrule
6000116200 & 2014-12-16 & 50.0  & Harrison \\
\midrule
\multicolumn{4}{c}{\textbf{\textit{Swift} XRT}} \\
\midrule
00080825002 & 2014-12-19 & 2.0 & Swift \\
\bottomrule
\end{tabular}
\begin{center}
\end{center}
\end{table}

\subsection{Archival Swift and NuSTAR Observations}
\label{sec:nustar_obs}
NGC~5005 was observed with \textit{NuSTAR} \citep{harrison_nuclear_2013} on December 16, 2014 (ObsID: 60001162002), with a total exposure time of approximately 50~ks. A short follow-up snapshot was obtained by \textit{Swift} \citep{gehrels_swift_2004} on December 19, 2014, with a total exposure of $\sim$2~ks. The \textit{NuSTAR} FPMA and FPMB data were analyzed in the 3–30~keV band, while the \textit{Swift}-XRT data were examined over the 0.5–4~keV range. Details of both observations are listed in Table~\ref{tab:xray_observations}.

\textit{NuSTAR} data reduction was performed using the \texttt{NUSTARDAS}\footnote{\url{https://heasarc.gsfc.nasa.gov/docs/nustar/analysis/nustar\_swguide.pdf}} software package with the latest calibration files available as of April 15, 2025. Science products were generated with the \texttt{nuproducts} task for both focal plane modules (FPMA and FPMB). The source spectrum was extracted from a circular region of radius $30''$ centered on the nucleus of NGC~5005. A background spectrum was extracted from a nearby source-free region of equal size. Spectra were grouped to ensure a minimum of five counts per bin.

The \textit{Swift}-XRT exposure was processed using the automated pipeline provided by the Space Science Data Center (SSDC) of the Italian Space Agency. We used their interactive interface to reduce, calibrate, and clean the dataset, and to extract both source and background spectra. The source region was defined as a circle with radius $20''$ centered on the AGN, while the background was extracted from an adjacent annulus with a width of $10''$, positioned to sample a blank region. The resulting spectrum was grouped to contain at least three counts per bin.

\section{Imaging Analysis and Results}
\label{sec:imaging_analysis_results}

\subsection{Modeling the Chandra PSF}
\label{sec:psf_model}

To accurately model the \textit{Chandra} point spread function (PSF), we simulated individual PSFs for each of the \textit{Chandra} exposures listed in Table~\ref{tab:xray_observations}. These simulations were carried out using \texttt{CHART}\footnote{\url{https://cxc.cfa.harvard.edu/ciao/PSFs/chart2/index.html}}, which accounts for the input source spectrum, off-axis angle, and aspect solution of each observation. The simulated ray traces were then projected onto the ACIS-S detector using \texttt{MARX}\footnote{\url{https://space.mit.edu/cxc/marx/}}, which incorporates parameters such as observation date, detector position, and exposure time.

To minimize pixelation effects in the final PSF model, we set the \texttt{ASPECTBLUR} parameter to 0.19, as recommended for ACIS-S data\footnote{\url{https://cxc.cfa.harvard.edu/ciao/why/aspectblur.html}}. Each individual \texttt{CHART} simulation was converted into an event file after projection. These simulated PSF event files were then combined using \texttt{dmmerge}\footnote{\url{https://cxc.cfa.harvard.edu/ciao/ahelp/dmmerge.html}} to create a single, composite PSF suitable for use in our imaging analysis.

All simulations were registered to a common astrometric frame using ObsID~25698 as the reference. Finally, the composite PSF was normalized to match the total source counts within a circular region of radius $r = 1.5''$ centered on the nucleus of NGC~5005. 

\begin{figure}
  \centering
  \includegraphics[width=.49\textwidth]{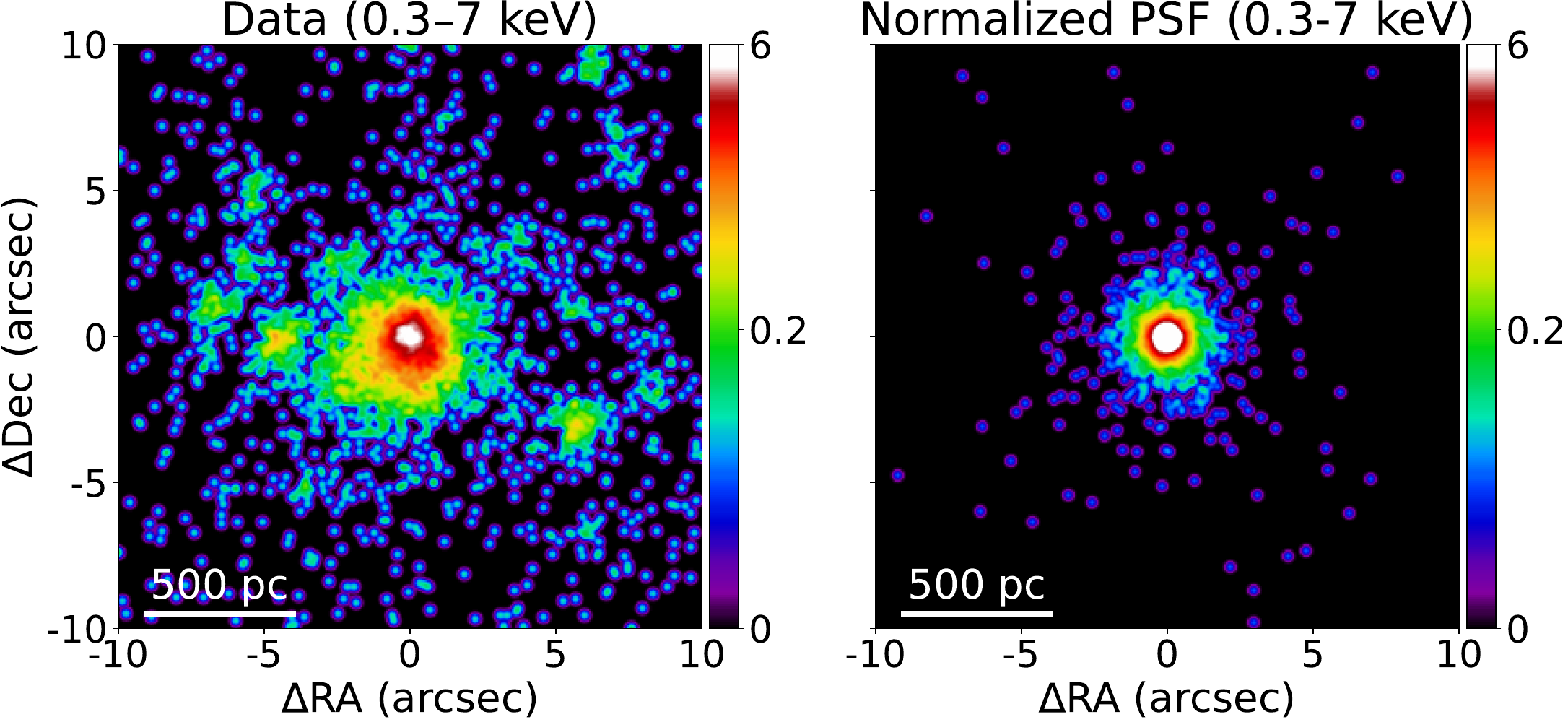}

\caption{\textit{Left:} Merged \textit{Chandra} 0.3–7~keV image of NGC~5005, as shown in Figure~\ref{fig:merged_image}. \textit{Right:} Simulated PSF generated with \texttt{MARX}, scaled to match the total number of counts within the central $1.5''$ radius. Both panels are displayed with the same color scale, binning (1/8 pixel), and Gaussian smoothing to enable direct comparison.}
 \label{fig:psf_comp} 
 \end{figure}

Figure~\ref{fig:psf_comp} displays a side-by-side comparison of the observed \textit{Chandra} image and the simulated, normalized PSF model in the 0.3–7~keV band. Both images use identical binning, smoothing, and color scales for direct visual comparison.

\subsection{Azimuthal and Radial Dependence of the Extended X-ray Emission}
\label{sec:extended_emission}

The merged \textit{Chandra} images presented in Figure~\ref{fig:merged_image} reveal a bright nuclear point source surrounded by diffuse circumnuclear X-ray emission extending out to $r \sim 5''$ (500~pc), with a roughly circular morphology. To isolate and characterize the extended component, we first identified point sources in the field using the \texttt{wavdetect}\footnote{\url{https://cxc.cfa.harvard.edu/ciao/ahelp/wavdetect.html}} algorithm with scale parameters of 1, 2, 4, 8, and 16 pixels.

Several point sources were detected across the field of view. These sources were excluded from both the spatial and spectral analysis. Their corresponding pixels were replaced using the \texttt{dmfilth}\footnote{\url{https://cxc.cfa.harvard.edu/ciao/ahelp/dmfilth.html}} tool, which interpolates pixel values from surrounding regions at comparable radial distances from the nucleus. The regions affected by the point sources were also masked from subsequent measurements.

After excluding the point sources, we detect $4752 \pm 69$ net counts above the background within a $5''$ radius from the centroid of the 0.3–7~keV emission (see Table\ref{tab:counts}). The X-ray emission in this region is dominated by soft photons, with only $943 \pm 31$ net counts detected at energies above 3~keV, indicating that the diffuse emission is primarily soft in nature.

\begin{figure}
  \centering
  \includegraphics[width=.49\textwidth]{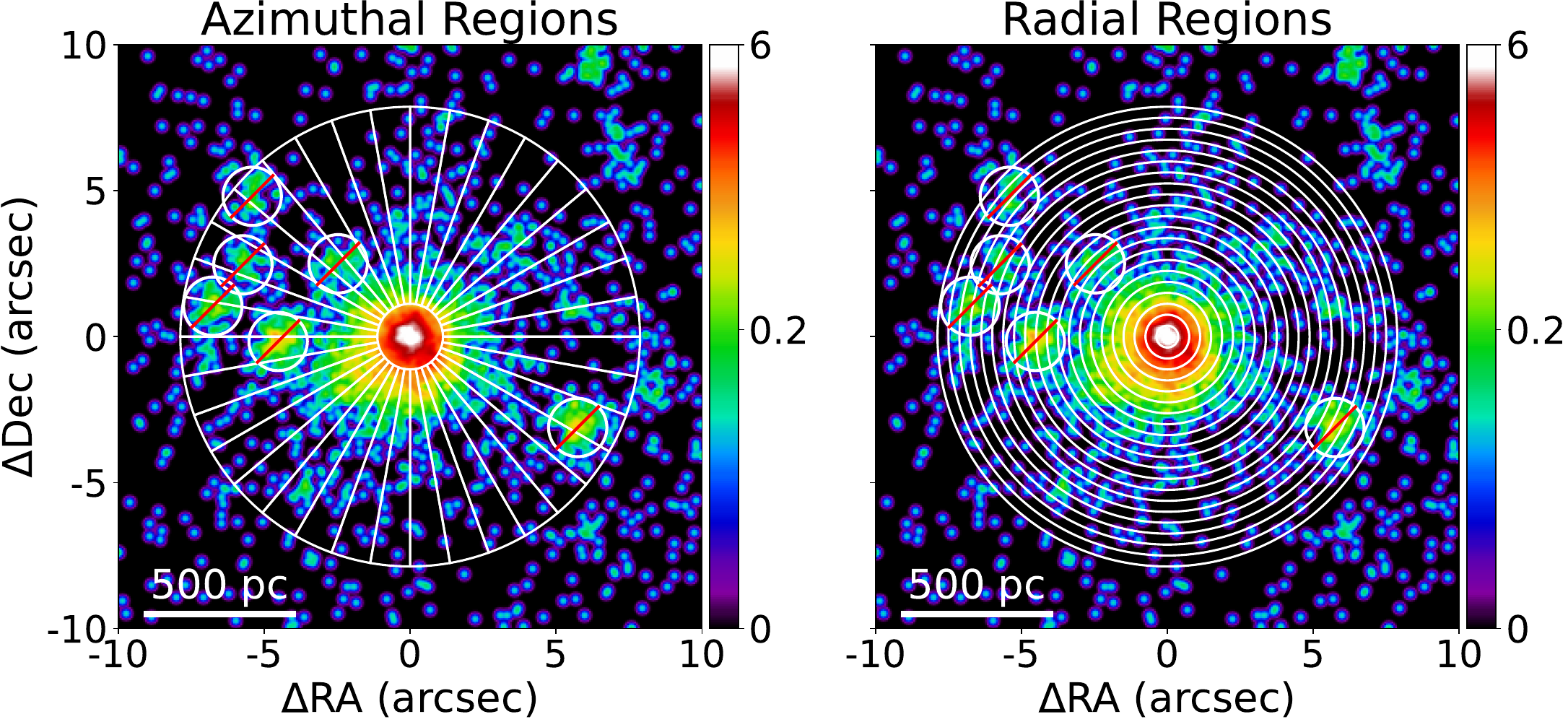}

\caption{\textit{Left panel:} \textit{Chandra} full-band image with overlaid pie-shaped extraction regions spanning radii of 1.5$''$–8$''$ and divided into 36 angular bins (10\degree\ each), used to extract azimuthal surface brightness profiles from the merged data and simulated PSF model}. \textit{Right panel:} Same image with 20 overlaid concentric annuli used to extract radial surface brightness profiles from a $r=$8$''$ region.
 \label{fig:extraction_regions_spatial} 
 \end{figure}

\begin{figure*}
  \centering
  \includegraphics[width=.9\textwidth]{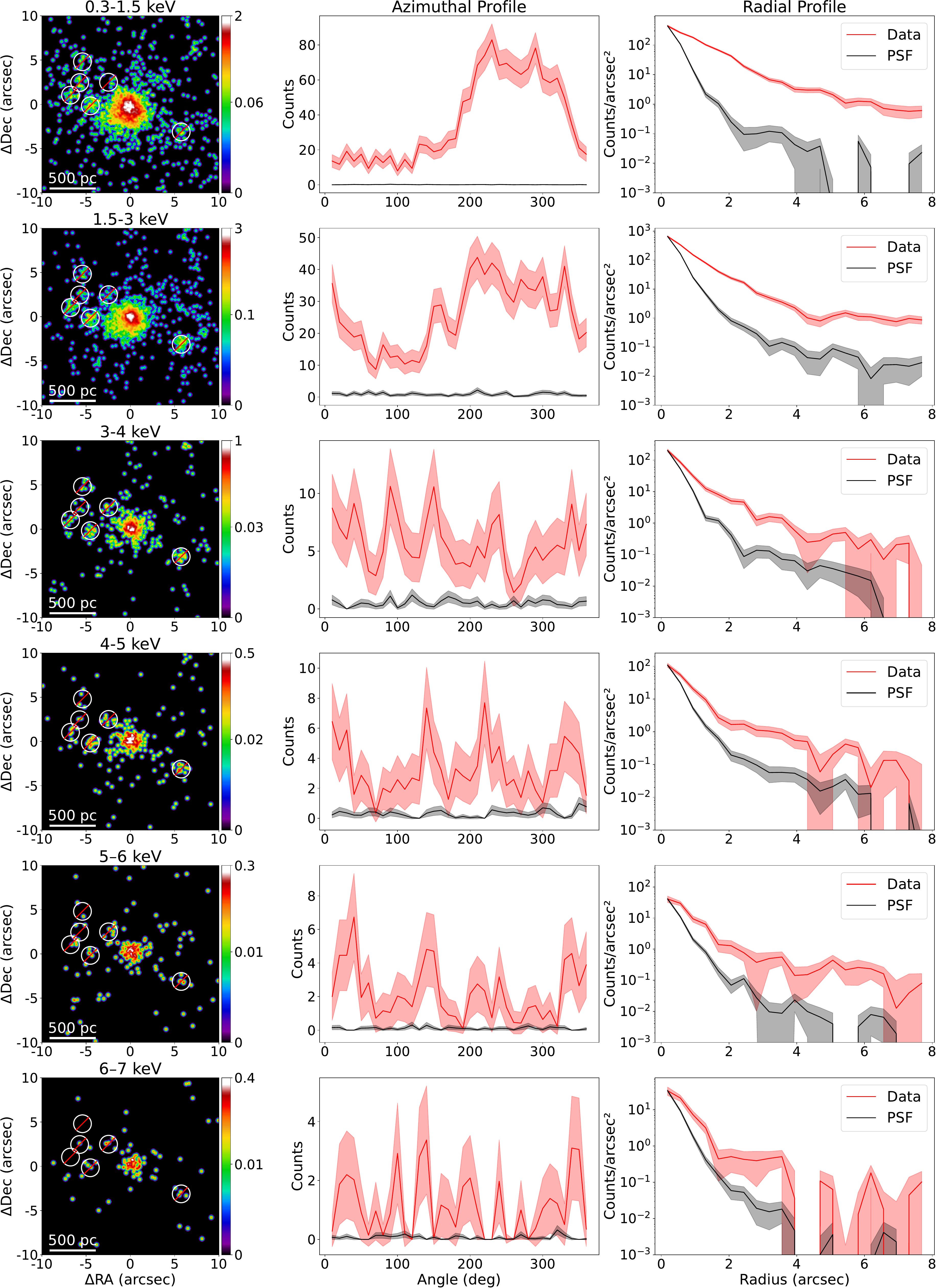}
\caption{Images (left column), azimuthal surface brightness profiles (middle column, 0\degree corresponds to West), and radial surface brightness profiles (right column) of NGC~5005 in different energy bands. The images were rebinned to 1/8 of the native ACIS pixel scale and smoothed with a 3-pixel Gaussian. The PSF in each band was normalized to match the source counts in the innermost region.}
 \label{fig:radial_profiles} 
 \end{figure*}

To investigate potential directional asymmetries in the extended emission and assess their dependence on photon energy, we constructed azimuthal surface brightness profiles in several narrow X-ray energy bands. These profiles were extracted from both the observed data and energy-matched PSF simulations. The extraction regions consisted of pie-shaped sectors extending from $r = 1.5''$ to $r = 8''$, centered on the centroid of the 0.3–7~keV image, and divided into $10^\circ$ angular bins (see Figure~\ref{fig:extraction_regions_spatial}, left). Point sources were excluded in all cases. The simulated PSF profiles were normalized to match the source counts within the innermost ($r<1.5''$) region. The middle panels of Figure~\ref{fig:radial_profiles} show a direct comparison between the observed data (shown in red) and PSF (in black) azimuthal profiles. 

To further quantify the extent of the diffuse emission, we constructed radial surface brightness profiles in each energy band and compared them with the corresponding PSF models. The right-hand panels of Figure~\ref{fig:radial_profiles} present the background-subtracted radial profiles from the merged dataset (shown in red), overlaid with the simulated PSF profiles (in black). All profiles were extracted from a circular region of radius $r = 8''$, centered on the centroid of the full-band image, and divided into 20 concentric annuli (see Figure~\ref{fig:extraction_regions_spatial}, right). The PSF simulations incorporate the exposure times of the individual observations (Table~\ref{tab:xray_observations}), so the model profiles include the expected level of statistical uncertainty.

\subsection{Spatial Properties of the Extended X-ray Emission}
\label{sec:spatial_properties}

As illustrated in Figure~\ref{fig:radial_profiles}, the soft X-ray emission (energies $<$3~keV) is the most spatially extended component. Radial surface brightness profiles in the 0.3–1.5~keV and 1.5–3~keV bands show emission reaching out to $r \sim 8''$ (800~pc) from the nuclear centroid. These profiles clearly exceed that of the simulated PSF, indicating the presence of diffuse, extended emission in these bands. The corresponding azimuthal profiles (middle column of Figure~\ref{fig:radial_profiles}) show that the soft X-ray emission is preferentially enhanced in the 180$^\circ$–360$^\circ$ range relative to the PSF, with excess emission extending toward the southeast (SE), south (S), and southwest (SW) directions.

We also analyzed the spatial properties of the emission in the 3–4~keV and 4–5~keV bands, which yield $2748 \pm 52$ and $1562 \pm 40$ net counts, respectively, within a circular region of radius $r = 8''$. After subtracting the PSF contribution, the residual extended emission in these bands corresponds to $671 \pm 26$ and $436 \pm 21$ counts, respectively (see Table~\ref{tab:counts}). The radial profiles in both bands show significant emission extending out to $r \sim 6''$–$7''$ (600–700~pc). However, unlike the soft bands, the azimuthal profiles in these higher energy ranges do not show strong directional asymmetries.

As the photon energy increases, the spatial extent of the X-ray emission decreases, while the azimuthal profiles remain relatively flat. The emission in the 5–6~keV band extends to $r \sim 7''$ (700~pc), whereas the 6–7~keV emission is confined to within $r \sim 4''$ (400~pc) of the nuclear centroid.

\begin{table}
\centering
\caption{Excess Net Counts Above Model PSF}
\label{tab:counts}
\begin{tabular}{ccc}
\toprule
\textbf{Band (keV)} & \textbf{$r<8''$} & \textbf{$1.5''<r<8''$} \\
\midrule
0.3-1.5 & $4537 \pm 67$ & $375 \pm 19$ \\
1.5-3.0 & $3639 \pm 60$ & $159 \pm 13$ \\
3.0-4.0 & $671 \pm 26$  & $75 \pm 9$   \\
4.0-5.0 & $436 \pm 21$  & $60 \pm 8$   \\
5.0-6.0 & $293 \pm 17$  & $26 \pm 5$   \\
6.0-7.0 & $172 \pm 13$  & $15 \pm 4$   \\
\bottomrule
\end{tabular}
\end{table}

\begin{figure}
  \centering
  \includegraphics[width=.45\textwidth]{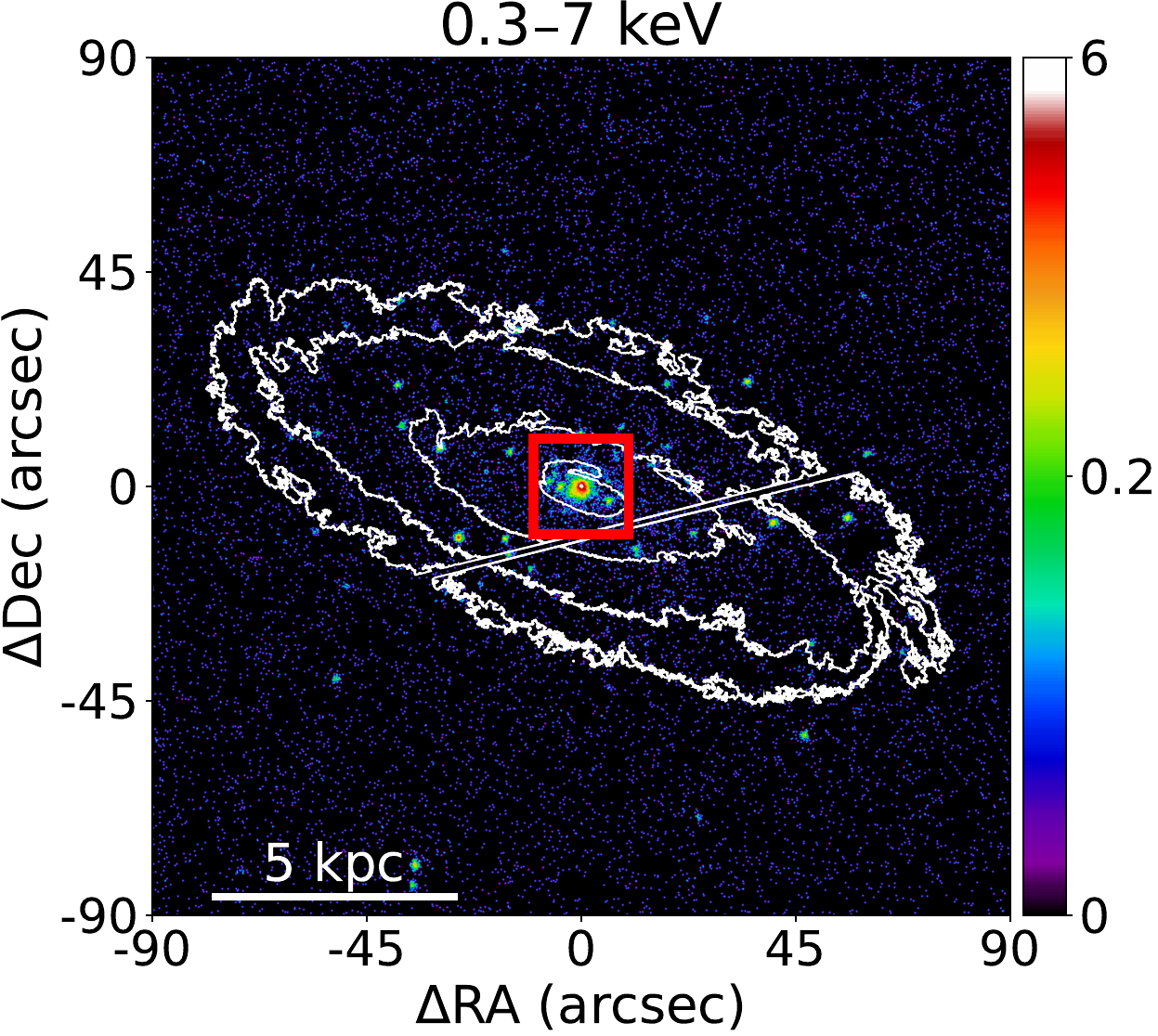}
\caption{Full-band \textit{Chandra} image of NGC~5005, rebinned to 1/8 of the native ACIS pixel scale. White contours show the \textit{HST} F814W wide-band image. The diffuse X-ray excess detected in the data (see radial profiles in Figure~\ref{fig:radial_profiles}) likely includes contributions from unresolved sources in the galaxy, such as X-ray binaries, supernova remnants, stellar sources, and hot interstellar medium. The red square has dimensions of 5$''\times5''$, as in Figure~\ref{fig:radial_profiles}, for comparison.}
 \label{fig:chandra_814w} 
 \end{figure}

Notably, the radial profiles at all energies (Figure~\ref{fig:radial_profiles}, right) do not decline to zero at large radii, even in the highest energy bands. This suggests the presence of a pervasive diffuse X-ray component throughout the galaxy. Supporting this interpretation, Figure~\ref{fig:chandra_814w} compares the full-band \textit{Chandra} image with contours from the \textit{HST} F814W wide-band continuum filter, revealing that the \textit{Chandra} field of view is largely filled by the host galaxy. The integrated X-ray emission from unresolved sources, such as X-ray binaries, supernova remnants, stellar populations, and hot ISM, likely contributes to the observed diffuse excess. A detailed analysis of this galaxy-wide emission component is beyond the scope of this work.

In summary, the radial surface brightness analysis indicates that the extended emission reaches out to $\sim$800~pc at energies below 3~keV, declines to $\sim$600–700~pc in the 3–6~keV range, and is restricted to $\sim$400~pc at 6–7~keV. While the soft-band emission shows directional enhancements toward the SE, S, and SW, the higher-energy emission appears more isotropic.

\section{Spectral Analysis and Results}
\label{sec:spectral_analysis_results}

\begin{figure*}
  \centering
  \includegraphics[width=\textwidth]{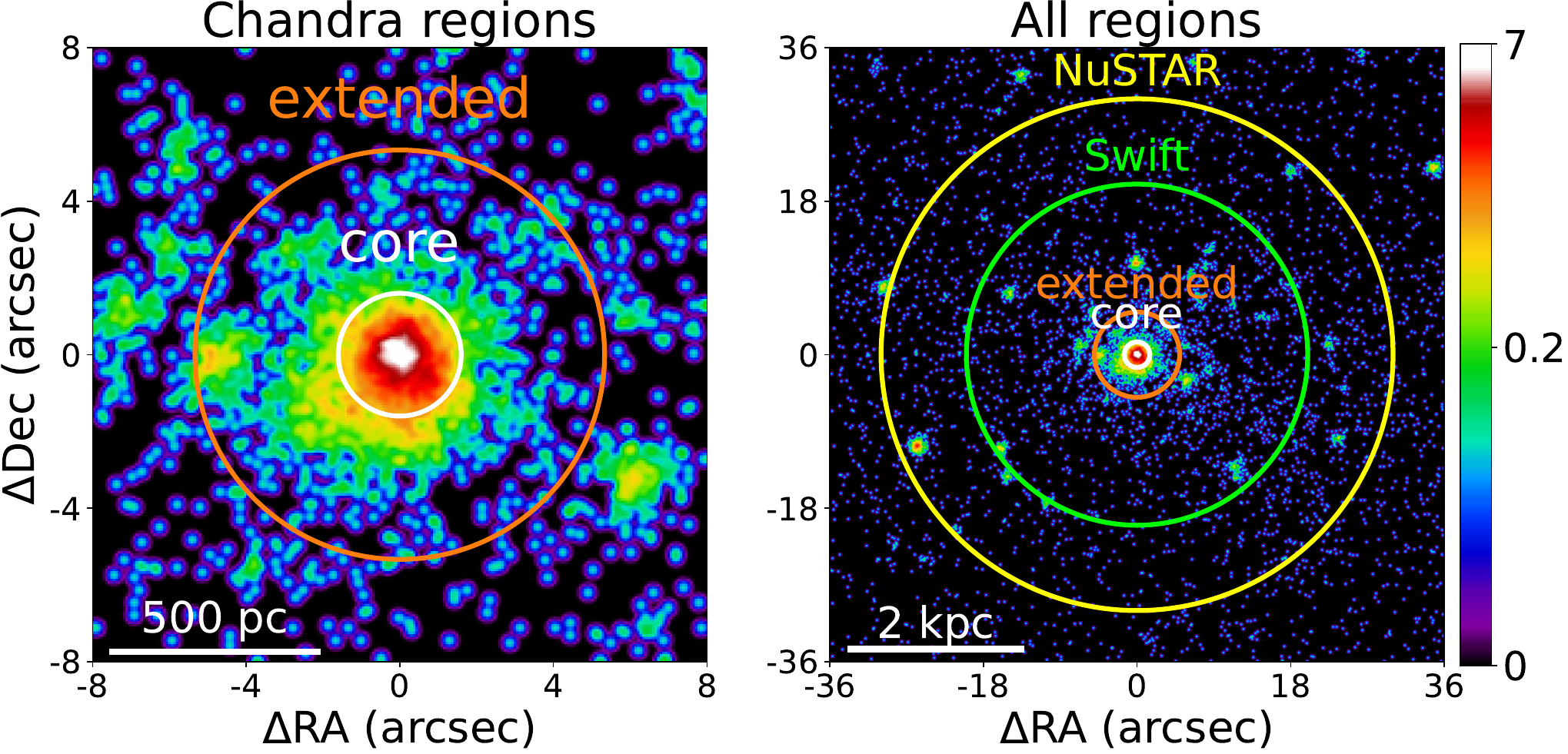}
\caption{\textit{Left:} \textit{Chandra} ACIS-S image of the 0.3–7~keV emission, rebinned to 1/8 of the native pixel scale and smoothed with a Gaussian kernel ($\sigma = 1.5$ pixels). The inner white circle ($r = 1.5''$) marks the \textit{Chandra} core extraction region, while the surrounding orange annulus ($1.5'' < r < 5''$) corresponds to the extended emission region analyzed in Section~\ref{sec:spectral_analysis_results}. \textit{Right:} Same as in the left panel, but with the outer green annulus indicating the \textit{Swift} extraction region ($r = 20''$), while the outermost yellow annulus corresponds to the \textit{NuSTAR} extraction region ($r = 30''$).}
 \label{fig:spec_extraction_regions} 
 \end{figure*}

Guided by the azimuthal and radial surface brightness profiles presented in Figure~\ref{fig:radial_profiles}, we extracted spectra from two distinct regions: (1) a circular region with radius $r = 1.5''$ (150~pc), centered on the centroid of the 0.3–7~keV emission, to isolate the nuclear and immediate circumnuclear emission (hereafter referred to as the \textit{core} spectrum); and (2) an annular region spanning $r = 1.5''$–$5''$ (150–500~pc), to isolate the extended circumnuclear emission (hereafter the \textit{extended} spectrum). These extraction regions are shown in the left panel of Figure~\ref{fig:spec_extraction_regions}.

Spectra for each of the sixteen individual \textit{Chandra} exposures (see Table~\ref{tab:xray_observations}) were extracted using the \texttt{CIAO} tool \texttt{specextract}\footnote{\url{https://cxc.cfa.harvard.edu/ciao/ahelp/specextract.html}}. A circular background region of radius $r = 10''$, free of point sources and located on the same CCD chip, was used for background subtraction. The resulting source and background spectra, as well as the corresponding response files, were combined across all observations using the \texttt{combine\_spectra}\footnote{\url{https://cxc.cfa.harvard.edu/ciao/ahelp/combine_spectra.html}} tool. The final spectra were grouped to contain a minimum of 10 counts per bin to ensure robust spectral fitting.

Spectral modeling was carried out using \texttt{XSPEC}\footnote{\url{https://heasarc.gsfc.nasa.gov/xanadu/xspec/}}, adopting physically motivated models appropriate for AGN and circumnuclear emission components, as detailed in the subsections below.

\subsection{Spectral Modeling}
\label{sec:spec_modeling}

To investigate the dominant emission mechanisms in each spatial region, we employed a physically motivated modeling approach incorporating both photoionized and collisionally ionized plasma components. Photoionized emission was modeled using precomputed grids generated with \texttt{Cloudy}\footnote{\url{http://www.nublado.org/}} \citep{ferland_2017_2017}, while thermal emission was modeled using the \texttt{APEC}\footnote{\url{http://www.atomdb.org/}} plasma code \citep{foster_updated_2012}.

Each extracted spectrum was initially fit with a single emission component. Additional components were introduced only when justified by statistically significant residuals and improvements in the fit statistic. Model adequacy was evaluated using both the C-statistic per degree of freedom (Cstat/d.o.f.) and visual inspection of residuals. In cases where Cstat/d.o.f.~$\sim$~1 but structured residuals remained, additional physically motivated components were incorporated.

\subsubsection{Photoionized and Thermal Emission Components}
\label{sec:photo_thermal_models}

The photoionized gas was modeled using additive emission components derived from \texttt{Cloudy} c13.01 grids. These models assume an input spectral energy distribution (SED) of the form $L_{\nu} \propto \nu^{-\alpha}$, with $\alpha = 0.5$ for $1\times10^{-4}$~eV~$< h\nu <$~13.6~eV and $\alpha = 1.0$ for 13.6~eV~$< h\nu <$~30~keV. Solar elemental abundances were adopted throughout. The model grid spans ionization parameters in the range log~$U$ = $-2$ to 4 (in steps of 0.25 dex) and hydrogen column densities from log~$N_{\rm H}$ = 19 to 23.5 (in steps of 0.1 dex). These grids were converted into \texttt{XSPEC}-compatible additive tables following the procedure described in \citet{porter_cloudyxspec_2006}.

Collisionally ionized thermal emission was modeled using \texttt{APEC}, assuming solar abundances. This component accounts for hot, diffuse gas possibly heated by AGN-driven winds or radio jets, or by stellar processes such as supernovae and star formation \citep[e.g.,][]{fabbiano_deep_2018, travascio_agn-host_2021, trindade_falcao_deep_2023}.

\subsection{Spectral Properties of the X-ray Emission}
\label{sec:spec_properties}

The results of the spectral modeling for each \textit{Chandra} spectra are summarized in Table~\ref{tab:spec_results}, with the extraction regions shown in the left panel of Figure~\ref{fig:spec_extraction_regions}. All fits include foreground Galactic absorption fixed at $N_{\rm H} = 1.18 \times 10^{20}$~cm$^{-2}$, as determined using the NASA HEASARC tool\footnote{\url{https://heasarc.gsfc.nasa.gov/}}.

\subsubsection{The Core Spectrum}
\label{sec:core_spec}
\begin{figure*}
  \centering
   \includegraphics[width=8.5cm]{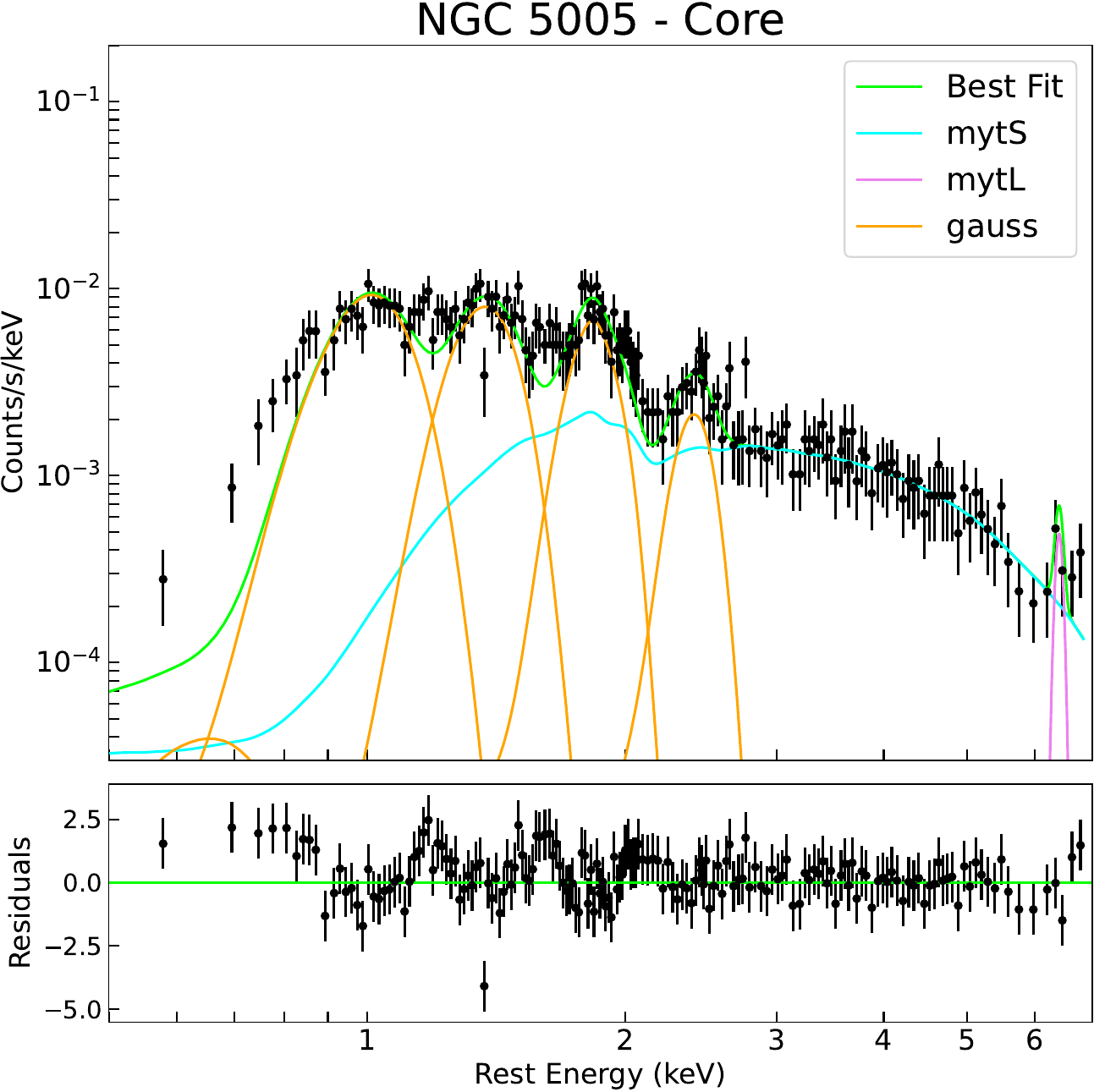}
   \includegraphics[width=8.5cm]{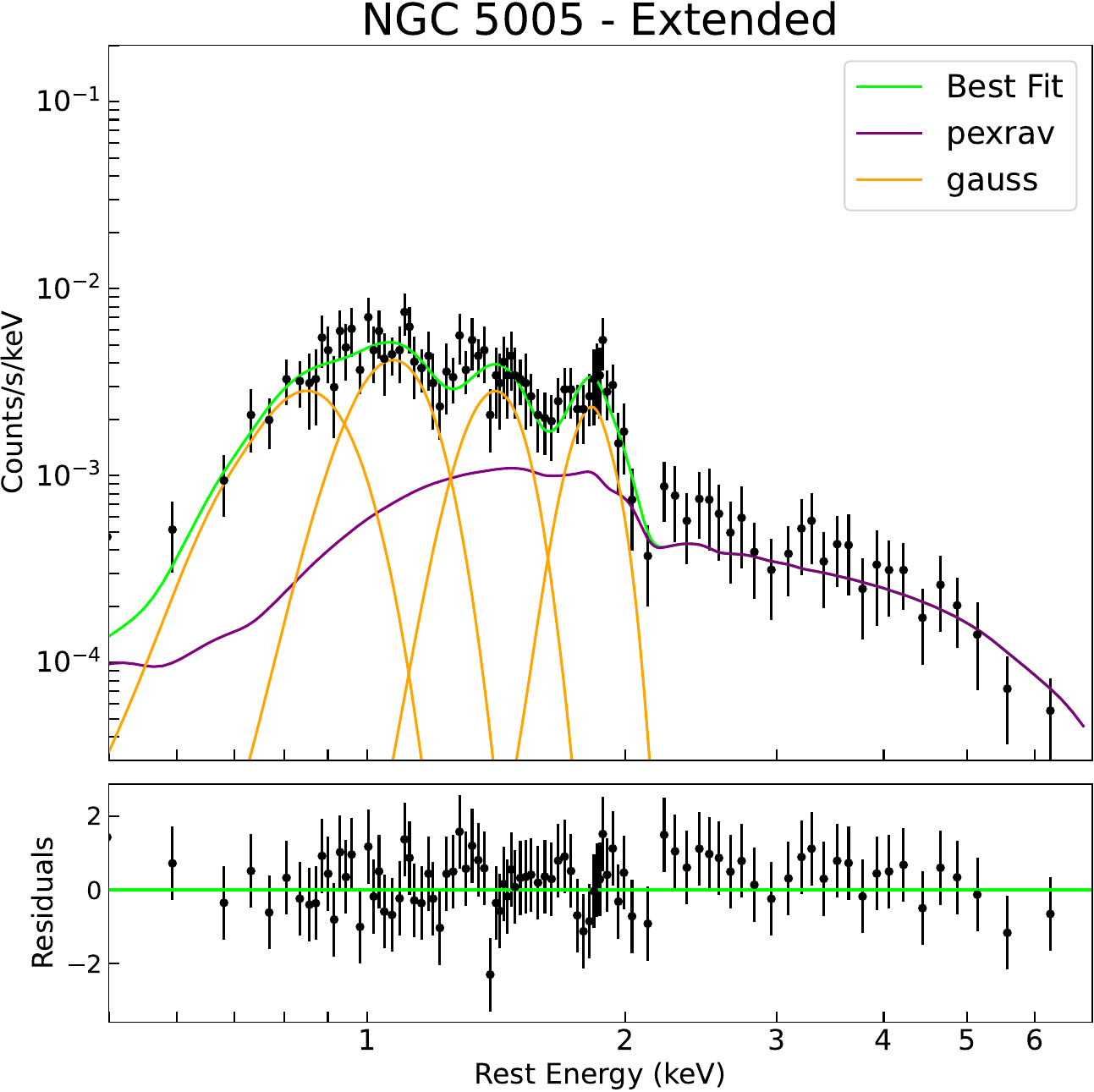}\\
   \includegraphics[width=8.5cm]{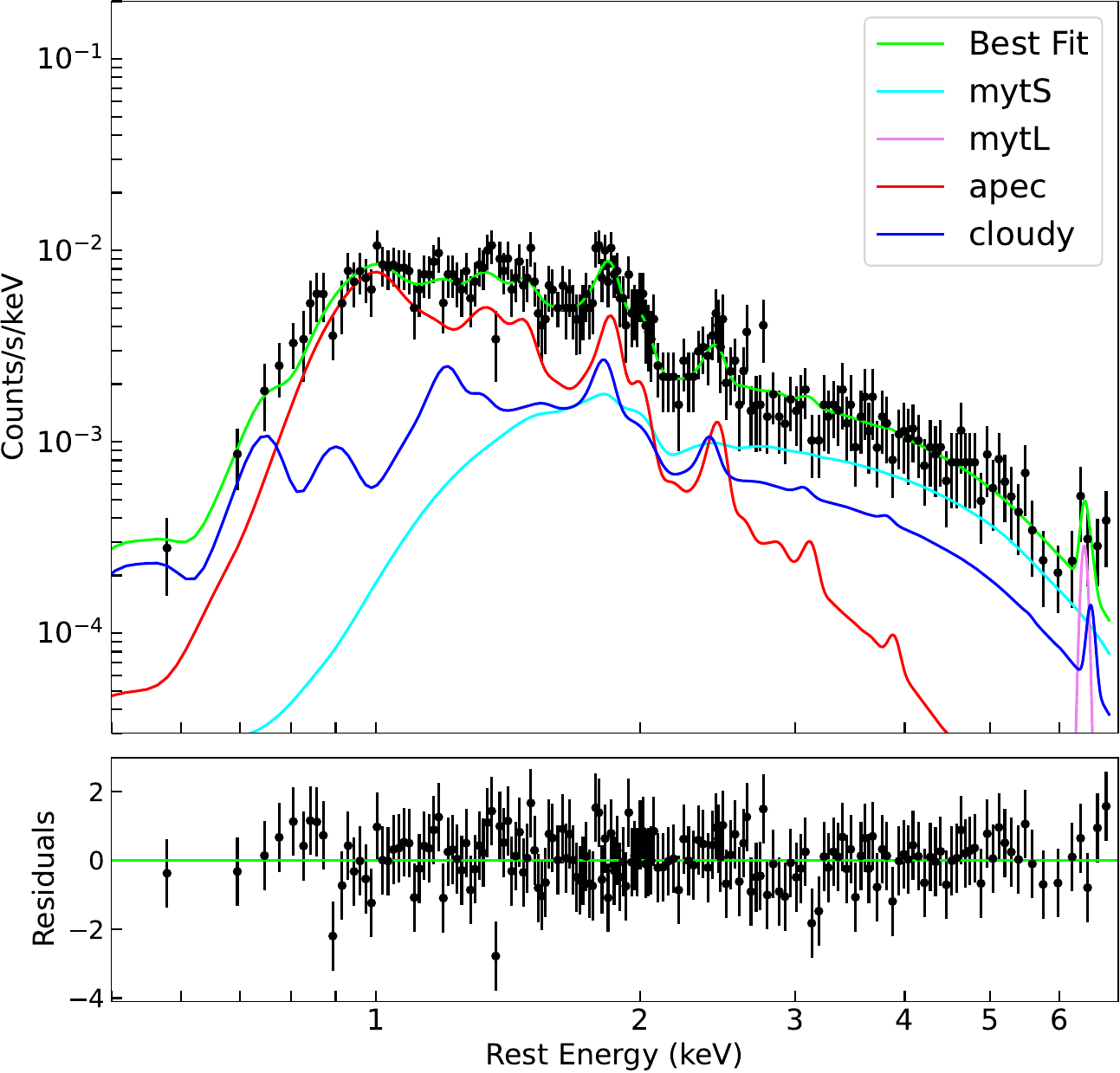}
   \includegraphics[width=8.5cm]{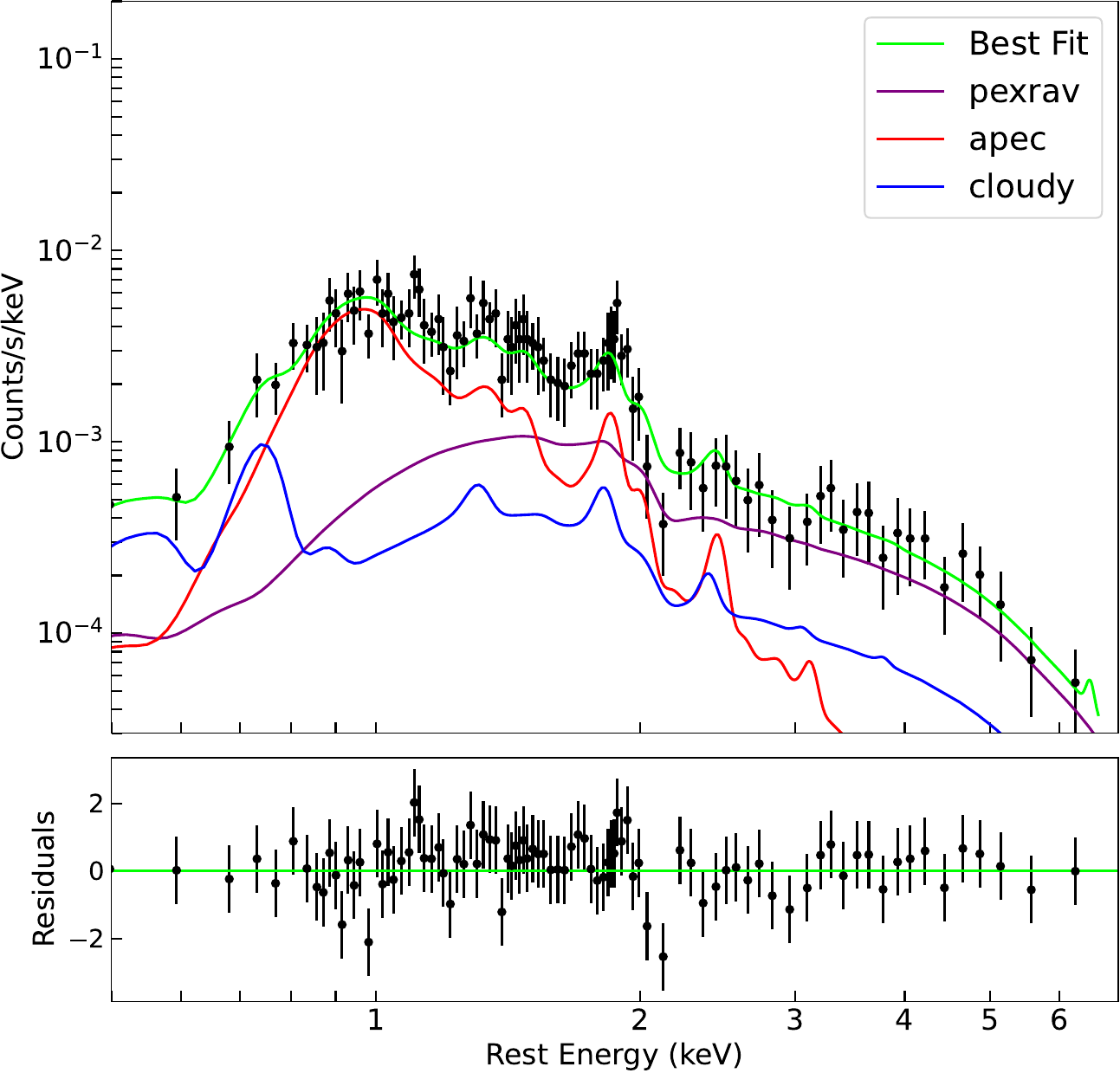}
\caption{\textit{Left panels:} X-ray spectrum of the core region ($r < 1.5''$), showing the best-fit phenomenological model} (top) and best-fit for a 1-photoionization + 1-thermal component model (bottom). The hard X-ray emission is modeled with a \texttt{MYTorusL} component for the Fe~K$\alpha$ line (pink) and a \texttt{MYTorusS} component for the scattered continuum (cyan). \textit{Right panel:} X-ray spectrum of the extended region ($1.5'' < r < 5''$), with the same model structure. The hard X-ray emission is modeled with an absorbed \texttt{pexrav} reflection component (purple).
 \label{fig:spectra} 
 \end{figure*}

The core spectrum (Figure~\ref{fig:spectra}, left) is dominated by soft X-ray emission below 3~keV, but also shows evidence of hard X-ray features, including a possible neutral Fe~K$\alpha$ line at $E_{\rm rest}=6.4$~keV. To model this high-energy component, we adopted the physically motivated torus model \texttt{MYTorus}\footnote{\url{https://www.mytorus.com/}} \citep{murphy_x-ray_2009}, which simulates X-ray reprocessing in a toroidal distribution of neutral material. Specifically, we included the scattered continuum (\texttt{mytS}, shown in cyan) and fluorescent line (\texttt{mytL}, shown in pink) components, to account for reprocessed hard X-ray emission and iron line fluorescence.

To identify regions dominated by line emission, we first performed a phenomenological fit to the soft X-ray band using several redshifted Gaussian components with fixed width $\sigma = 0.1$~keV, matching the spectral resolution of the ACIS-S detector. Line centroids and normalizations were allowed to vary freely. This model also included foreground absorption via \texttt{tbabs}. The results of this fit are shown in the top-left panel of Figure~\ref{fig:spectra}. We emphasize that this approach is not intended for deriving physical quantities of the gas, especially below 1.5~keV where spectral blending is significant at the resolution of ACIS.

This analysis reveals four prominent emission features in the core spectrum, with fitted centroid energies of $E_{\rm rest} = 0.94 \pm 0.01$~keV, $1.35 \pm 0.02$~keV, $1.84 \pm 0.02$~keV, and $2.40 \pm 0.03$~keV. These likely correspond to blended line complexes of Ne~IX+Ne~X, Mg~XI+Mg~XII, Si~XIII+Si~XIV, and S~XV, respectively \citep[e.g.,][]{trindade_falcao_deep_2023}.

We then proceed to fit the full core spectrum using a physically motivated model that combines photoionized gas (modeled with \texttt{Cloudy}) and collisionally ionized plasma (modeled with \texttt{APEC}). The photon index for both \texttt{mytS} and \texttt{mytL} components was fixed at $\Gamma = 1.8$, and their column densities were linked but allowed to vary. The inclination angle was fixed at 0$\degree$. The best-fit model (Cstat/d.o.f. = 1.04; shown in lime green in Figure~\ref{fig:spectra}) yields a line-of-sight column density of $N_{\rm H} = (3.7 \pm 0.3) \times 10^{21}$~cm$^{-2}$, consistent with previous studies \citep{gonzalez-martin_x-ray_2009, younes_study_2012}.

The best-fit model (Figure~\ref{fig:spectra}, bottom-left panel) indicates that the soft X-ray emission in the core is best described by a combination of a low-ionization photoionized component (log~$U = -0.11 \pm 0.11$, log~$N_{\rm H} = 21.00 \pm 0.22$~cm$^{-2}$; shown in blue) and a thermal plasma with $kT = 0.90 \pm 0.05$~keV (shown in red). Table~\ref{tab:spec_results} reports the corresponding model luminosities for each component in the 0.3–7~keV band. The observed and intrinsic 2–10~keV X-ray luminosities derived from the model are $L_{\rm obs} = (2.67 \pm 0.46) \times 10^{39}$~erg~s$^{-1}$ and $L_{\rm int} = (2.77 \pm 0.46) \times 10^{39}$~erg~s$^{-1}$, respectively.

\subsubsection{The Extended Spectrum}
\label{sec:ext_spec}

As shown in Figure~\ref{fig:radial_profiles}, extended X-ray emission is detected out to $r \sim 800$~pc, particularly at energies below 3~keV. In obscured AGNs, such soft X-ray emission is commonly associated with [O~III]$\lambda$5007 and is typically attributed to AGN photoionization, though contributions from collisionally ionized plasma may also be present \citep[e.g.,][]{fabbiano_interaction_2024}. 

The spectrum extracted from the extended region of NGC~5005 is shown in the right panels of Figure~\ref{fig:spectra}. To model the hard continuum, we included a reflection component using \texttt{pexrav} (shown in purple), assuming a power-law slope of $\Gamma = 1.8$, an exponential cutoff energy of $E_{\rm fold} = 1,000$~keV, and an inclination cosine of $\cos(i) = 0.5$.

As with the core spectrum, we first applied a phenomenological model to identify dominant emission lines in the soft band. This model included several redshifted Gaussian components (in orange) with fixed widths of $\sigma = 0.1$~keV and variable centroid energies and normalizations. The result of this fit is shown in the top-right panel of Figure~\ref{fig:spectra}. The extended spectrum exhibits strong emission features at $E_{\rm rest} = 0.77 \pm 0.03$~keV, $1.03 \pm 0.03$~keV, $1.39 \pm 0.03$~keV, and $1.83 \pm 0.03$~keV. These features likely correspond to blended emission from O~VIII+Fe~XVII, Ne~IX+Ne~X, Mg~XI+Mg~XII, and Si~XIII+Si~XIV, respectively.

We then fit the full extended spectrum using a combination of photoionized and thermal plasma components, as described in Section~\ref{sec:photo_thermal_models}. The best-fit model (Cstat/d.o.f.~$= 1.13$; shown in lime green in Figure~\ref{fig:spectra}) includes a low-ionization photoionized component with log~$U = -0.76 \pm 0.35$ and log~$N_{\rm H} = 19.37 \pm 0.24$~cm$^{-2}$ (shown in blue), along with a collisionally ionized thermal component with $kT = 0.89 \pm 0.05$~keV (shown in red).

\begin{table*}
\centering
\caption{Best-fit Spectral Components and Parameters for NGC~5005}
\label{tab:spec_results}
\begin{tabular}{lllcc}
\toprule
Component & Physical Component & Parameter & Value & $L_{\rm 0.3-7~keV}$ \\
\midrule
\multicolumn{5}{c}{\textbf{Core Region (Cstat/d.o.f. = 1.04)}} \\
\midrule
\texttt{TBabs} & Foreground absorption & $N_{\rm H}$ & $0.37 \pm 0.03$ & – \\
\texttt{apec} & Collisionally-ionized gas & $kT$ & $0.90 \pm 0.05$ & – \\
\texttt{apec} & Collisionally-ionized gas & norm & $(7.37 \pm 1.02)\times10^{-5}$ & $39.68 \pm 0.03$  \\
\texttt{CLOUDY} & Photoionized gas & log $U$ & $-0.11 \pm 0.11$ & –  \\
\texttt{CLOUDY} & Photoionized gas & log $N_{\rm H}$ & $21.00 \pm 0.22$ & –  \\
\texttt{CLOUDY} & Photoionized gas & norm & $(1.10 \pm 1.25)\times10^{-14}$ & $39.84 \pm 0.10$  \\
\texttt{mytL} & Fluorescent lines (torus) & $N_{\rm H}$ & $<$2.00 & –  \\
\texttt{mytL} & Fluorescent lines (torus) & norm & $(2.58 \pm 1.15)\times10^{-3}$ & $37.81 \pm 0.20$  \\
\texttt{mytS} & Scattered AGN continuum (torus) & norm & $(8.37 \pm 1.02)\times10^{-3}$ & $39.14 \pm 0.06$  \\
\midrule
\multicolumn{5}{c}{\textbf{Extended Region (Cstat/d.o.f. = 1.13)}} \\
\midrule
\texttt{apec} & Collisionally-ionized plasma & $kT$ & $0.89 \pm 0.05$ & –  \\
\texttt{apec} & Collisionally-ionized plasma & norm & $(1.89 \pm 0.56)\times10^{-5}$ & $39.09 \pm 0.03$  \\
\texttt{CLOUDY} & Photoionized gas & log $U$ & $-0.76 \pm 0.35$ & –  \\
\texttt{CLOUDY} & Photoionized gas & log $N_{\rm H}$  & $19.37 \pm 0.24$ & –  \\
\texttt{CLOUDY} & Photoionized gas & norm & $(4.20 \pm 2.00)\times10^{-13}$ & $38.92 \pm 0.13$  \\
\texttt{pexrav} & Reflected AGN continuum (disk/slab) & norm & $(6.11 \pm 1.32)\times10^{-6}$ & $38.84 \pm 0.07$  \\
\midrule
\multicolumn{5}{c}{\textbf{\textit{NuSTAR–Swift} (Cstat/d.o.f. = 0.88)}} \\
\midrule
\texttt{TBabs} & Foreground absorption & $N_{\rm H}$  & $0.37^{\dagger}$ & –  \\
\texttt{apec} & Collisionally-ionized plasma & $kT$ & $0.90^{\dagger}$ & - \\ 
\texttt{apec} & Collisionally-ionized plasma & norm & $<1.2 \times10^{-4}$ & $39.83 \pm 0.14$  \\
\texttt{CLOUDY} & Photoionized gas & log $U$ & $-0.11^{\dagger}$ & –  \\ 
\texttt{CLOUDY} & Photoionized gas & log $N_{\rm H}$  & $21.00^{\dagger}$& –  \\ 
\texttt{CLOUDY} & Photoionized gas & norm & $(5.6 \pm 2.6)\times10^{-14}$ & $<39.78$ \\
\texttt{mytL} & Fluorescent lines (torus) & $N_{\rm H}$ & $<$2.00$^{\dagger}$ & –  \\
\texttt{mytL} & Fluorescent lines (torus) & norm & $(1.17 \pm 0.07)\times10^{-2}$ & $38.64 \pm 0.20$  \\
\texttt{mytS} & Scattered AGN continuum (torus) & norm & $<4.1 \times10^{-2}$ & $39.65 \pm 0.25$  \\
\bottomrule
\end{tabular}
\begin{center}
\textbf{Note:} $N_{\rm H}$ is in units of $10^{22}$~cm$^{-2}$, $kT$ in keV, and luminosities are in log erg~s$^{-1}$. "norm" refers to the \texttt{XSPEC} normalization of each component. \texttt{CLOUDY} and \texttt{apec} represent photoionized and thermal gas, respectively. \texttt{mytS} and \texttt{mytL} are reprocessed continuum and line emission from a toroidal AGN structure. The $\dagger$ symbols identify the parameters that were frozen during the fitting process, see Section~\ref{sec:NuSTAR_spec_analysis} for details.
\end{center}
\end{table*}

\subsection{NuSTAR-Swift Spectroscopic Analysis}
\label{sec:NuSTAR_spec_analysis}

\begin{figure}
  \centering
   \includegraphics[width=8.5cm]{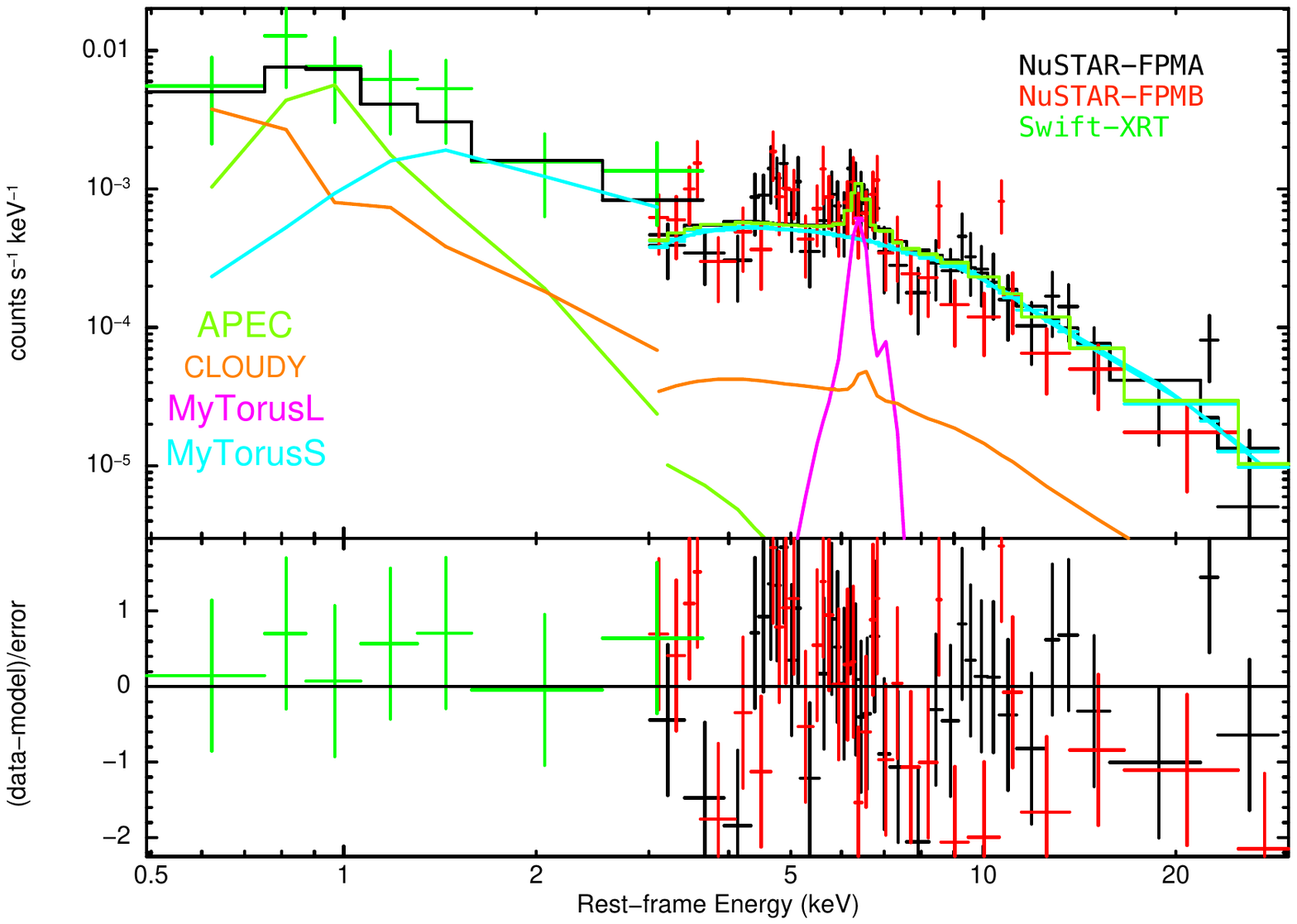}
\caption{Joint \textit{NuSTAR–Swift} spectra of NGC~5005 and best-fit model, with residuals shown below. The individual model components are indicated.}
 \label{fig:nuunfold} 
 \end{figure}

To complement our \textit{Chandra} analysis, we also examined archival \textit{NuSTAR} and \textit{Swift} observations of NGC~5005 (Table~\ref{tab:xray_observations}, Section~\ref{sec:nustar_obs}). The extraction regions for the \textit{NuSTAR} and \textit{Swift} data are shown in the right panel of Figure~\ref{fig:spec_extraction_regions}.

Figure~\ref{fig:nuunfold} displays the unfolded (i.e., corrected for instrumental response) \textit{NuSTAR-Swift} spectra, with \textit{NuSTAR} FPMA and FPMB shown in black and red, respectively, and \textit{Swift} XRT in green. The \textit{Swift} data show mild excess emission at soft X-rays, while both \textit{NuSTAR} spectra reveal the presence of a Fe~K$\alpha$ emission line at $E_{\rm rest} \sim 6.4$~keV. The \textit{Swift} XRT spectrum extends up to $\sim$~4keV, while the \textit{NuSTAR} FPMA and FPMB spectra were analyzed from 3 to 30~keV, above which the background dominates.

To account for cross-calibration and extraction region differences between instruments (FPMA/B and XRT), we included a multiplicative \texttt{constant}\footnote{\url{https://heasarc.gsfc.nasa.gov/xanadu/xspec/manual/node249.html}} component in \texttt{XSPEC}. The FPMA spectrum was treated as the reference, while the normalization constant for FPMB and XRT were allowed to vary. The two {\it NuSTAR} spectra are found to be compatible within a few percent, and the \textit{Swift} spectrum is consistent with the \textit{NuSTAR} FPMA data within $\sim$15\%, as expected in a joint analysis \citep{Madsen2015}.

We first fit a simple absorbed power-law model, fixing the Galactic column density at $N_{\rm H} = 1.18 \times 10^{20}$~cm$^{-2}$. This yielded a photon index of $\Gamma = 2.02 \pm 0.13$ and a fit statistic of Cstat/d.o.f.$= 0.90$. Adding a narrow Gaussian line with a fixed width of $\sigma = 50$~eV (below the resolution of \textit{NuSTAR}) improved the fit significantly ($\Delta$Cstat/$\Delta$d.o.f.$= -15/-2$), resulting in a line centroid at $E_{\rm rest} = 6.3 \pm 0.1$~keV and normalization of $(1.7 \pm 0.8) \times 10^{-6}$~photons~cm$^{-2}$~s$^{-1}$. This model yields an upper limit for the absorbing column of $N_{H}<1.9\times10^{21}$~cm$^{-2}$, along with observed and intrinsic 2–10~keV source luminosities of $L^{\rm po}_{\rm obs} = (7.34 \pm 0.52) \times 10^{39}$~erg~s$^{-1}$ and $L^{\rm po}_{\rm int} = (8.54 \pm 0.60) \times 10^{39}$~erg~s$^{-1}$, respectively.

Next, we tested whether the best-fit model derived from the \textit{Chandra} core region (Table~\ref{tab:spec_results}) could describe the joint \textit{NuSTAR–Swift} data. All parameters were fixed to their \textit{Chandra} best-fit values, except for their normalizations. This model achieved a good fit (Cstat/d.o.f.$= 0.88$) and is shown in Figure~\ref{fig:nuunfold}; the best-fit parameters and 0.3–7~keV component luminosities are listed in Table~\ref{tab:spec_results}.

The derived intrinsic luminosities for individual model components from this fit are slightly higher than in the \textit{Chandra} core spectrum, but consistent within $\sim$2$\sigma$. This difference is likely due to minor contributions from unresolved point sources within the larger apertures used for the extraction of the \textit{NuSTAR} and \textit{Swift} spectra (see Figure~\ref{fig:spec_extraction_regions}, right). Additionally, the \texttt{mytL} component normalization and luminosity are slightly enhanced compared to the \textit{Chandra} core spectrum, which may reflect the higher effective area of the \textit{NuSTAR} detectors in the 6–7~keV energy range.

The final best-fit model (shown in Figure~\ref{fig:nuunfold}) yields observed and intrinsic 2–10~keV luminosities of $L_{\rm obs} = (7.62 \pm 0.45) \times 10^{39}$~erg~s$^{-1}$ and $L_{\rm int} = (7.91 \pm 0.44) \times 10^{39}$~erg~s$^{-1}$, consistent with those obtained with the simpler absorbed power-law model.

\section{Discussion}
\label{sec:discussion}

It is well established that AGNs are powered by the accretion of material onto supermassive black holes (SMBHs) located at the centers of galaxies \citep{rees_black_1984}. The diversity observed in AGN emission spectra is often attributed to differences in nuclear obscuration and the observer’s line of sight, as described by the AGN unification model (see \citealt{netzer_revisiting_2015} for a review). AGNs also play a key role in galaxy evolution by injecting energy and momentum into the host galaxy through radiation, relativistic jets, and outflows, thereby influencing star formation processes \citep[e.g.,][]{sazonov_quasars_2004}.

High-resolution observations of nearby AGNs with \textit{Chandra} have enabled studies of AGN-host galaxy interactions on spatial scales ranging from several kiloparsecs down to a few hundred parsecs. In particular, imaging and spectroscopic observations of local Seyfert galaxies (e.g., NGC 4151, \citealt{wang_deep_2011-2, wang_deep_2011-1, wang_deep_2011}; Mrk 573, \citealt{paggi_cheers_2012}; ESO 428-G014, \citealt{fabbiano_deep_2018, fabbiano_deep_2019}; NGC~5728, \citealt{trindade_falcao_deep_2023, trindade_falcao_discovery_2024, trindade_falcao_deep_2024}) have revealed the co-existence of AGN-photoionized clouds and gas shock-heated by radio jets. Our investigation of NGC~5005 builds on this line of work. Below, we explore the physical conditions in the core and extended regions using constraints from our analysis. The spatial and spectral properties of the emission help constrain the composition of the ISM (Section~\ref{sec:physical_em_models}) and provide insights into the nature of the central source (Section~\ref{sec:true_nature}).

\subsection{Physical Properties of the Circumnuclear X-ray Gas}
\label{sec:physical_em_models}

The 0.3–7~keV spectra of both the core and extended \textit{Chandra} regions of NGC~5005 reveal a relatively flat continuum and multiple partially blended emission lines at soft energies ($<$3~keV; Figure~\ref{fig:spectra}). Thanks to the deep cumulative \textit{Chandra} exposure, we are able to clearly distinguish the inner core region ($r\lesssim1.5''$, $\lesssim$150~pc) from the more extended region ($r=1.5''$–5$''$, 150–500~pc). 

The core spectrum is best described by a \texttt{MYTorus} reflection model (scattered + fluorescent line emission), with a fixed photon index of $\Gamma=1.8$, which dominates the emission above 3~keV. At softer energies, the emission is well reproduced by a two-component model: a photoionized component and a collisionally ionized thermal plasma.

The primary spectral difference between the core and extended regions lies in the need for the scattered continuum and Fe~K$\alpha$ fluorescence components in the core spectrum. In contrast, the extended emission is adequately modeled with a single \texttt{pexrav} reflection continuum at high energies.

\begin{figure*}
  \centering
   \includegraphics[width=\textwidth]{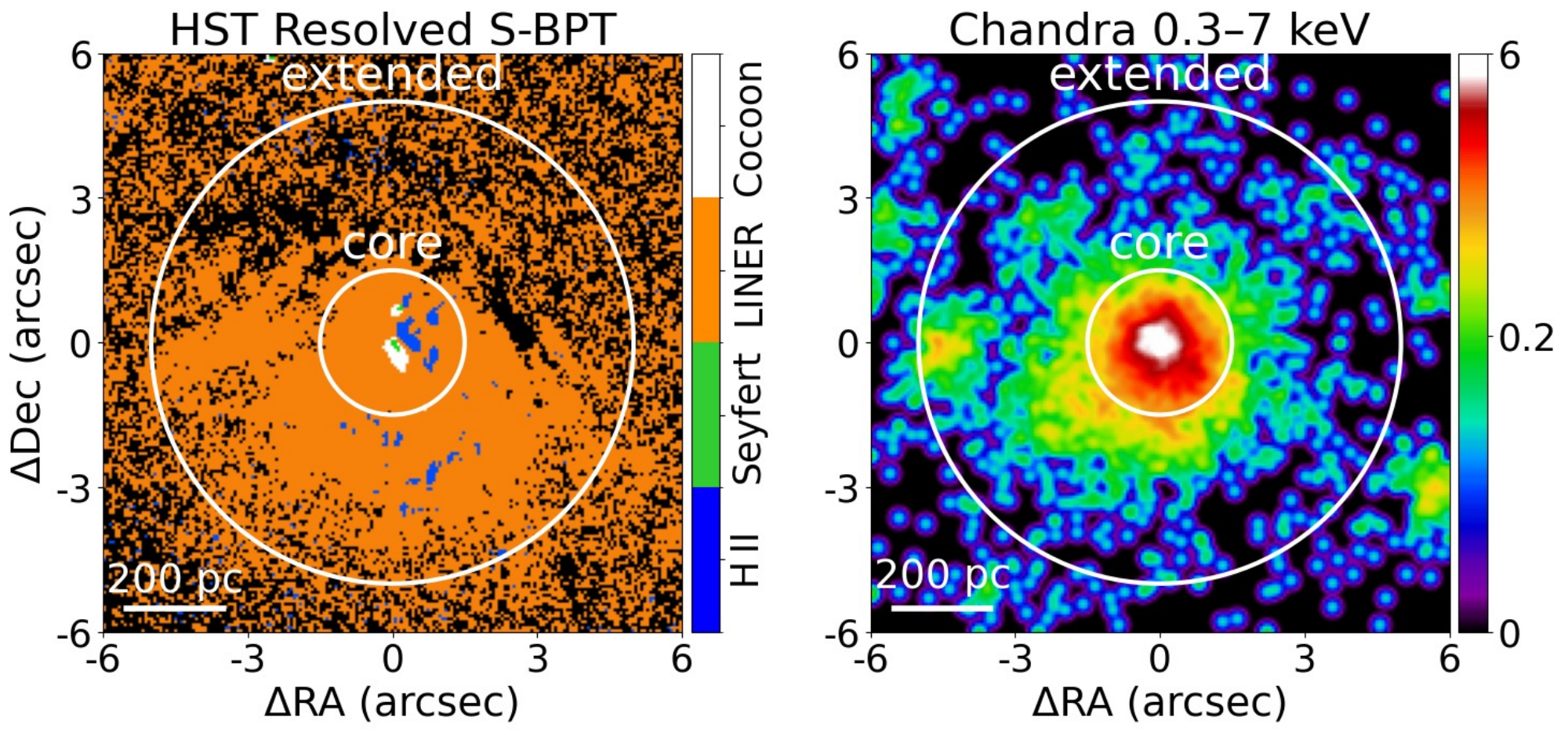}
\caption{\textit{Left panel:} Excitation map constructed from \textit{HST} narrow-band imaging, showing spatially resolved emission-line diagnostics. Seyfert-like regions are shown in green, H~II-like in blue, LINER-like in orange, and cocoon-like in white. \textit{Right panel:} \textit{Chandra} 0.3–7~keV image, binned to 1/8 of the native ACIS pixel scale. In both panels, the inner white circle marks the nuclear (core) spectral extraction region, while the outer annulus delineates the extended extraction region.}
 \label{fig:bpt_comp} 
 \end{figure*}

In Figure~\ref{fig:bpt_comp}, we compare the spatial distribution of optical excitation derived from spatially-resolved \textit{HST} data (left; from \citealt{trindade_falcao_mapping_2025}) with the 0.3–7~keV \textit{Chandra} image of NGC~5005 (right). The inner circle marks the core extraction region ($r<1.5''$), and the outer circle marks the extended extraction annulus ($1.5''<r<5''$). In the excitation map (left), orange denotes LINER-like excitation, blue H~II (star-forming)-like, green Seyfert-like, and white cocoon-like emission, corresponding to higher excitation LINER-like values in the S-BPT diagram \citep{trindade_falcao_mapping_2025}. Black pixels represent regions below the 3$\sigma$ detection threshold.

Within the core region, we detect a Seyfert-like nucleus (green) coinciding with the hard X-ray peak, surrounded by cocoon-like (white) and H~II-like (blue) regions. The Seyfert-like nucleus likely drives the need for a photoionized X-ray component in the core, while the cocoon and H~II-like regions are consistent with thermal plasma emission. In the extended region, LINER-like excitation (orange) dominates, with localized H~II-like contribution. In \citet{trindade_falcao_mapping_2025}, we argued that this extended LINER-like emission may be powered by a low-luminosity AGN, consistent with the presence of a photoionized component in the extended X-ray spectrum. The thermal component may originate from shock-heated gas in the NLR, as proposed by \citet{terao_near-infrared_2016}.

 \begin{figure}
  \centering
   \includegraphics[width=.5\textwidth]{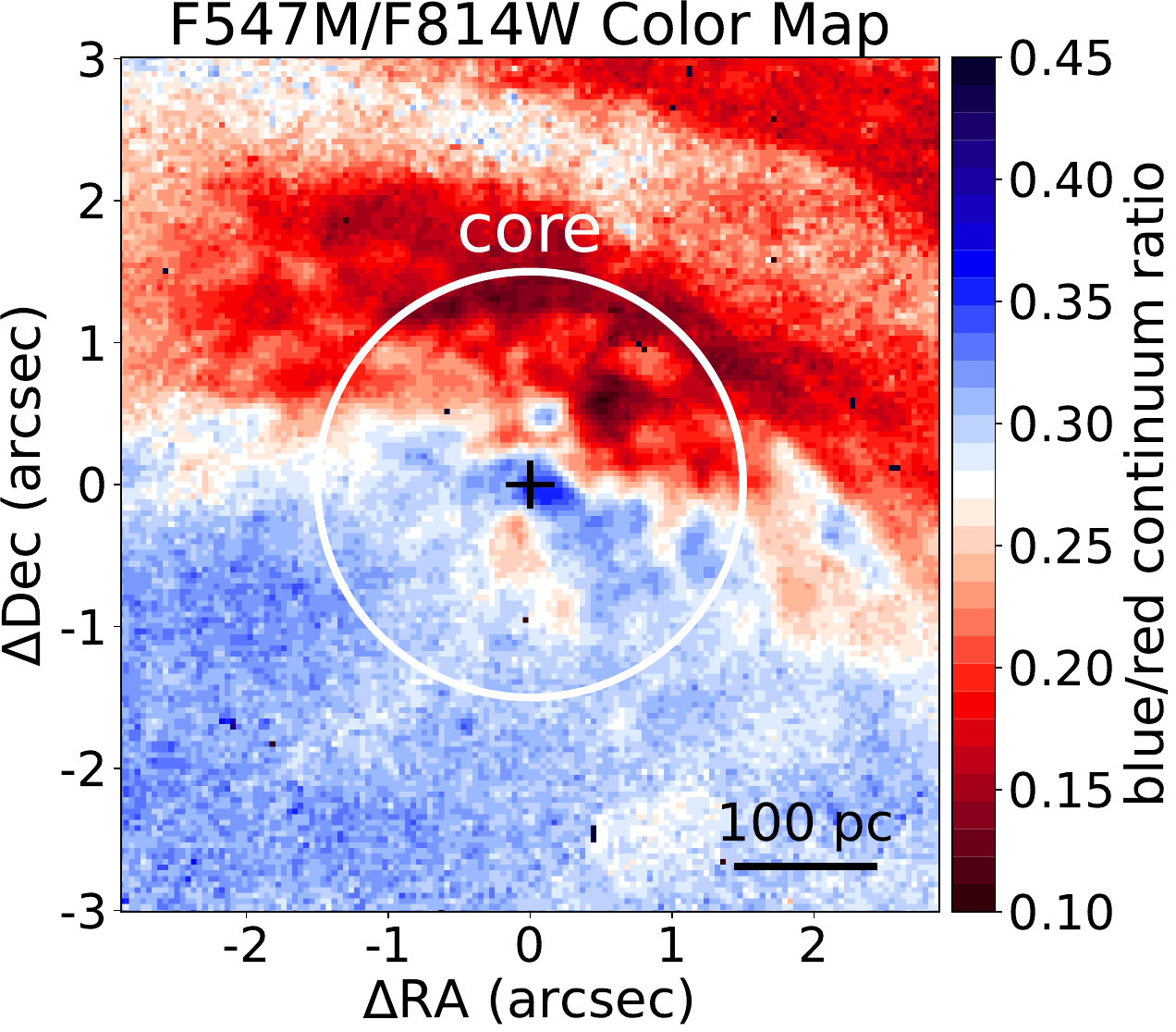}
\caption{Color map derived from the \textit{HST} F547M/F814W ratio, tracing the blue-to-red continuum variation across the central region of NGC~5005 \citep{trindade_falcao_mapping_2025}. Dark red regions indicate areas with significant dust extinction. The circle has a radius of 1.5$''$ and marks the \textit{Chandra} core extraction region, while the black cross marks the position of the \textit{Chandra} 0.3-7~keV centroid.}
 \label{fig:dust_map} 
 \end{figure}

Our spectral modeling (Section~\ref{sec:spectral_analysis_results}) confirms the presence of thermal components in both the core and extended regions, with $kT \sim 0.90$~keV (Figure~\ref{fig:spectra}; Table~\ref{tab:spec_results}). This temperature corresponds to a shock velocity of $v \sim 870$~km~s$^{-1}$, assuming $v^2_{\rm shock} = 16kT_{\rm shock}/3\mu$ (where $k$ is the Boltzmann constant and $\mu$ is the mean molecular mass of a fully ionized gas; e.g., \citealt{travascio_agn-host_2021}). Such shocks could arise from jet–ISM interactions or starbursts-driven winds, as suggested in \citealt{trindade_falcao_mapping_2025}.

We estimate the physical parameters of the thermal gas assuming a simple spherical geometry for the core region and a spherical shell for the extended region. The resulting quantities, as a function of the volume filling factor $\eta$, are summarized in Table~\ref{tab:apec_properties}. Derived electron densities range from $0.073\eta^{-1/2}$ to $0.86\eta^{-1/2}$~cm$^{-3}$, consistent with typical ISM values in AGN host galaxies \citep[e.g.,][]{baldi_chemical_2006, fabbiano_deep_2018, trindade_falcao_deep_2024}.

\begin{table}
\centering
\caption{Thermal Gas Parameters for NGC~5005}
\label{tab:apec_properties}
\begin{tabular}{lcc}
\toprule
\textbf{Parameter} & \textbf{Core} & \textbf{Extended} \\
\midrule
$kT$ (keV) & $0.90 \pm 0.04$ & $0.89 \pm 0.05$ \\
Volume $V$ ($10^{63}$ cm$^3$) & 0.4 & 15 \\
Electron density $n_e$ ($10^{-2}~\eta^{-1/2}$ cm$^{-3}$) & $86.1 \pm 6.7$ & $7.3 \pm 0.3$ \\
Gas mass $M_{\rm gas}$ ($10^{5}~\eta^{1/2}~M_\odot$) & $2.5 \pm 0.2$ & $7.6 \pm 2.6$ \\
Thermal energy $E_{\rm th}$ ($10^{53}~\eta^{1/2}$ erg) & $7.7 \pm 1.0$ & $23 \pm 1.8$ \\
Cooling time $t_{\rm cool}$ ($10^{6}~\eta^{-1/2}$ yr) & $1.9 \pm 0.5$ & $22 \pm 4.2$ \\
\bottomrule
\end{tabular}
\end{table}

For the core, the thermal energy content is $E_{\rm th} = 7.7\eta^{1/2} \times 10^{53}$~erg, with a radiative cooling time of $t_{\rm cool} = 1.9\eta^{-1/2} \times 10^{6}$~yr. Maintaining this temperature would require an energy injection rate of $E \sim 1.3 \times 10^{40}$~erg~s$^{-1}$. Alternatively, the same energy could be supplied by $\sim$7,730$\eta^{1/2}$ supernovae, assuming a 10\% thermalization efficiency \citep{chevalier_wind_1985, thornton_energy_1998}, corresponding to a supernova rate of $\sim$4.0$\eta^{1/2} \times 10^{-3}$~yr$^{-1}$.

To evaluate whether a recent supernova burst could contribute to the observed thermal X-ray emission in the core, we examined the optical color distribution using \textit{HST} imaging. Specifically, we revisited the F547M/F814W color map from \citet[][Fig.~1]{trindade_falcao_mapping_2025}, which traces blue-to-red continuum variations. Figure~\ref{fig:dust_map} displays the inner region of this color map, with the X-ray core extraction region overlaid. Regions strongly affected by dust extinction appear as dark red. While some reddening is expected due to dust lanes, the diffuse emission within $r=1.5''$ (white circle) is slightly bluer (mean F547M/F814W ratio $\sim$0.28). Furthermore, \textit{HST}/STIS spectroscopy of NGC~5005 \citep[e.g.,][]{cazzoli_optical_2018} reveals Balmer absorption features characteristic of a post-starburst phase \citep[e.g.,][]{dressler_spectroscopy_1983}, hinting at the presence of a young or intermediate-age stellar population in the nuclear region.

In the extended region, we estimate a thermal energy of $E_{\rm th} = 2.3\eta^{1/2} \times 10^{54}$~erg, which requires an energy injection rate of $E \sim 3.3 \times 10^{39}$~erg~s$^{-1}$ to maintain its observed temperature.

\subsection{True Nature of the Central Source}
 \label{sec:true_nature}

\begin{figure}
  \centering
   \includegraphics[width=.5\textwidth]{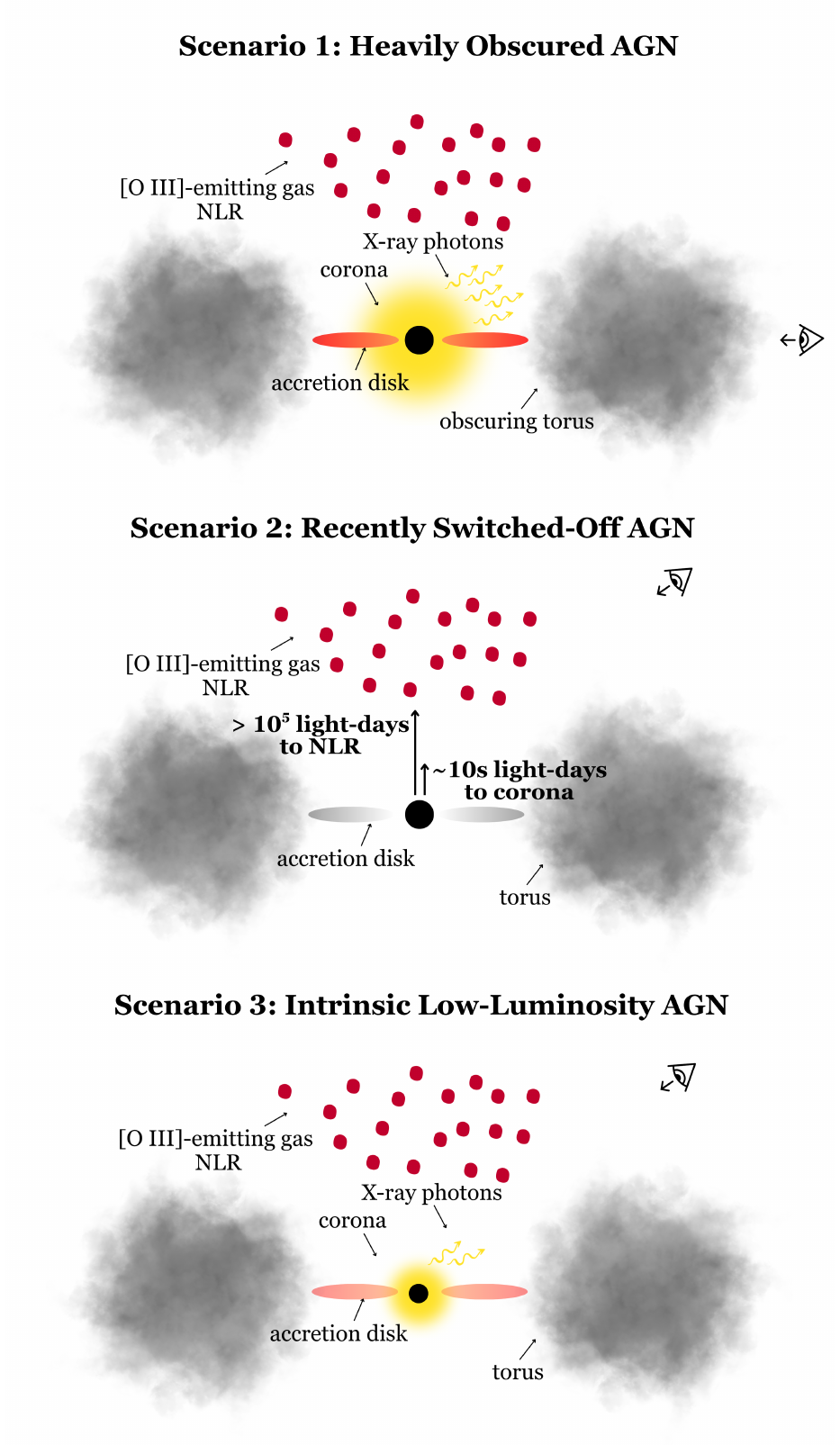}
\caption{Summary of possible scenarios for the nature of the central engine in NGC~5005. The panels illustrate three hypotheses: (1) a heavily obscured AGN, in which strong absorption could suppress the intrinsic X-ray luminosity; (2) a recently switched-off AGN, where the central engine has faded but [O~III] emission from the extended narrow-line region persists; and (3) an intrinsically low-luminosity AGN, where the central source is faint due to low accretion activity. All three scenarios are considered to explain the low observed X-ray luminosity derived from our spectral modeling.}
 \label{fig:cartoon} 
 \end{figure}

The low observed 2–10~keV luminosities derived from both the \textit{Chandra} core spectrum and the joint \textit{NuSTAR–Swift} fit suggests one of three possible scenarios for the central engine in NGC~5005: (1) the AGN is heavily obscured, (2) the AGN has recently switched off, or (3) the AGN is intrinsically faint. These scenarios are illustrated schematically in Figure~\ref{fig:cartoon}.

In the sections that follow, we evaluate each possibility in light of our spectral modeling results, aiming to better constrain the physical nature of the nuclear source in NGC~5005.

\subsubsection{A Heavily Obscured AGN?}
\label{sec:obscuration}

From the \textit{Chandra} core spectrum (Section~\ref{sec:core_spec}), we derive an observed 2–10~keV luminosity of $L_{\rm obs} = (2.67 \pm 0.46) \times 10^{39}$~erg~s$^{-1}$ and an absorbing column density of $N_{\rm H} = (3.7 \pm 0.3) \times 10^{21}$~cm$^{-2}$. This results in an intrinsic luminosity nearly identical to the observed one, $L_{\rm int} = (2.77 \pm 0.46) \times 10^{39}$~erg~s$^{-1}$. 

The joint \textit{NuSTAR–Swift} fit (Section~\ref{sec:NuSTAR_spec_analysis}) yields slightly higher luminosities, $L_{\rm obs} = (7.62 \pm 0.45) \times 10^{39}$~erg~s$^{-1}$ and $L_{\rm int} = (7.91 \pm 0.44) \times 10^{39}$~erg~s$^{-1}$, with a similarly low column density of $N_{\rm H} < 1.9 \times 10^{21}$~cm$^{-2}$. The modest increase in luminosity likely reflects contamination from nearby sources within the larger  extraction regions (Figure~\ref{fig:spec_extraction_regions}, right).

Both fits consistently indicate low absorption, with column densities on the order of $10^{21}$~cm$^{-2}$, strongly suggesting that the nucleus of NGC~5005 is not heavily obscured. Further supporting this conclusion, the equivalent width of the neutral Fe~K$\alpha$ line derived from the joint \textit{NuSTAR–Swift} fit is $EW = 287 \pm 100$~eV, well below the $\sim$1~keV typically expected for Compton-thick AGN.

\subsubsection{A Recently Switched-off AGN?}
\label{sec:faded?}

In AGNs, the hard X-ray continuum originating from the corona can vary on short timescales (days to months), reflecting the compact size of the emitting region \citep[e.g.,][]{wilkins_caught_2014, ricci_destruction_2020}. In contrast, the [O~III] emission is produced in the more extended NLR and responds to changes in the central engine on much longer timescales (hundreds to thousands of years). As a result, when an AGN switches off, the coronal X-ray emission fades rapidly, while the [O~III] luminosity remains elevated for some time (see Figure~\ref{fig:cartoon}). A recently faded AGN is thus expected to exhibit a large discrepancy between its measured X-ray luminosity and that inferred from [O~III] emission.

In \citet{trindade_falcao_mapping_2025}, we reported an extinction-corrected [O~III] luminosity of $L_{\rm [O~III]} = 1.2 \times 10^{39}$~erg~s$^{-1}$. Using the $L_{\rm [O~III]}$–$L_{\rm X}$ relation from \citet{lamastra_bolometric_2009}, this corresponds to an expected observed X-ray luminosity of $L_{\rm obs} = 1.3 \times 10^{40}$~erg~s$^{-1}$, consistent within uncertainties with the value derived from our joint \textit{NuSTAR–Swift} fit. 

This agreement suggests that the central engine of NGC~5005 has not recently switched off. Instead, the evidence points to the presence of an intrinsically low-luminosity AGN.

\subsection{An Intrinsically Low-Luminosity AGN}
\label{sec:LLAGN}

Using the observed 2–10~keV X-ray luminosity derived from the joint \textit{NuSTAR–Swift} spectral fit, and applying the bolometric correction from \citet{duras_universal_2020}, we estimate a bolometric luminosity of $L_{\rm bol} = (1.2 \pm 0.1) \times 10^{41}$~erg~s$^{-1}$. This value is consistent with the [O~III]-based bolometric luminosity estimate from \citet{trindade_falcao_mapping_2025}, $L_{\rm bol} = (1.0 \pm 0.7) \times 10^{41}$~erg~s$^{-1}$.

Adopting a black hole mass of $\log M_{\rm BH} = 8.27~M_{\odot}$ from \citet{beifiori_upper_2009}, we calculate an Eddington ratio of $\lambda_{\rm Edd} = L_{\rm bol}/L_{\rm Edd} \sim 5.1 \times 10^{-6}$, firmly placing NGC~5005 within the low-luminosity AGN (LLAGN) regime.

\begin{figure}
  \centering
  \includegraphics[width=.45\textwidth]{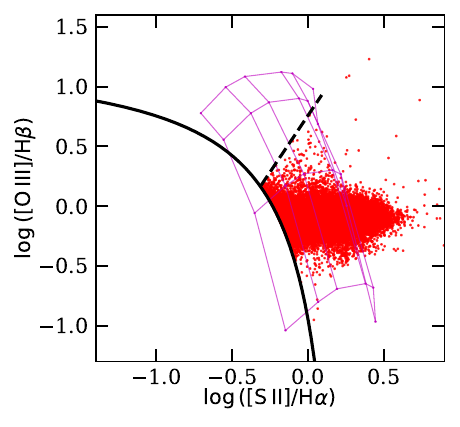}
\caption{Comparison between the AGN photoionization model tailored to NGC~5005 (magenta grid) and the observed LINER-classified pixels (red dots) on the S-BPT diagram \citep{trindade_falcao_mapping_2025}. The model grid includes lines of constant metallicity (from left to right: 12 + log(O/H) = 7.99, 8.28, 8.43, 8.54, 8.70, 8.80, 8.96, 9.09) and lines of constant ionization parameter (from bottom to top: log$U$ = –4.0, –3.5, –3.0, –2.5).}
 \label{fig:sed_models} 
\end{figure}

To further test this interpretation, we compare the emission-line fluxes predicted by photoionization models to the observed emission-line ratios on the spatially resolved S-BPT diagram from \citet{trindade_falcao_mapping_2025} (see Figure~\ref{fig:sed_models}). The photoionization model adopts a configuration similar to that of \citet{zhu_new_2023}, but utilizes an AGN SED tailored to the broadband properties of NGC~5005. 

As shown in Figure~\ref{fig:sed_models}, the model successfully reproduces most of the LINER-like emission observed in this galaxy. The only exceptions are pixels with unusually high [S~II]/H$\alpha$ ratios (log([S~II]/H$\alpha$) $>$ 0.4), which are rare among SDSS galaxies \citep[e.g.,][]{kewley_host_2006}. These may reflect an underestimation of the H$\alpha$ flux due to contamination from [N~II] in the \textit{HST} narrow-band filter used in the original map \citep{trindade_falcao_mapping_2025}. 

Overall, the consistency between photoionization model predictions and the observed emission-line ratios strongly supports the interpretation that NGC~5005 hosts an intrinsically faint, low-accretion-rate AGN.

\subsection{A Low Accretion Rate AGN Broad-Line Region}
\label{sec:unobscured_view_blr}

Theoretical models predict that the broad-line region (BLR) disappears at low accretion rates or bolometric luminosities. In the model proposed by \citet{nicastro_broad_2000}, a disk wind capable of launching the BLR can only form if the accretion rate exceeds $\dot{m} \gtrsim 10^{-3}$. Similarly, \citet{baskin_dust_2017} suggest that dusty photoionized gas can suppress BLR formation below $\dot{m} < 6.5 \times 10^{-4}\, (Z/Z_{\odot})^{-1}$, particularly in low-metallicity environments. 

Using $\dot{m} = L_{\rm bol}/(\eta c^{2})$, and adopting $L_{\rm bol} = 1.2 \times 10^{41}~\text{erg~s}^{-1}$ (Section~\ref{sec:LLAGN}) and a standard radiative efficiency of $\eta = 0.1$, we obtain $\dot{m} \approx 2.1 \times 10^{-5}~M_{\odot}~\text{yr}^{-1}$. This inferred accretion rate lies well below both thresholds, indicating that NGC~5005 is unlikely to host a BLR.

This expectation is seemingly at odds with some observational claims of a broad H$\alpha$ emission line in the optical spectrum \citep[e.g.,][]{balmaverde_hst_2014, cazzoli_optical_2018}. Constraints from \textit{HST} spectroscopy place an upper limit on the luminosity of this broad H$\alpha$ component, $L_{\rm H\alpha,\,b} < 4.2 \times 10^{38}$~erg~s$^{-1}$ \citep{balmaverde_hst_2014}. This upper limit is consistent (within uncertainties) with predictions from empirical type 1 AGN scaling relations \citep{stern_type_2012, stern_type_2012-1}, which yield an expected luminosity range of $(2.2 \pm 1.5) \times 10^{38} < L_{\rm H\alpha,\,b} < (6.0 \pm 1.0) \times 10^{38}$~erg~s$^{-1}$.

Similar challenges to the standard accretion paradigm have been reported in other systems. For example, \citet{bianchi_hst_2019} identified a relativistically broadened H$\alpha$ line in NGC~3147, interpreted as emission from a thin accretion disk, despite its low Eddington ratio of $\lambda_{\rm Edd} \sim 10^{-4}$. Such detections challenge the canonical view that AGNs transition to an optically thin, geometrically thick accretion flow at low accretion rates \citep[e.g.,][]{blandford_fate_1999}, and suggest that standard thin disks may persist even in the low-luminosity regime.

\section{Summary and Conclusions}
\label{sec:conclusions}

We present results from a deep 255~ks \textit{Chandra} ACIS observation of NGC~5005, a LINER-dominated galaxy previously reported to host broad H$\alpha$ emission. The data reveal extended X-ray emission with a roughly symmetric morphology, tracing diffuse structure out to $\sim$800~pc at soft energies ($<$3~keV), decreasing to $\sim$400~pc at higher energies (6–7~keV). A prominent central peak in the 0.3–7~keV image, spatially coincident with a Seyfert-like nucleus identified in \textit{HST} imaging, is interpreted as nuclear emission.

Spectral analysis of the nuclear ($r < 150$~pc) and extended ($150 < r < 500$~pc) regions reveals strong soft X-ray lines, consistent with contributions from both photoionized and collisionally ionized gas. In the nuclear region, the best-fit model includes a low-ionization ($\log U = -0.11$), low-column density ($\log N_{\rm H} = 21$~cm$^{-2}$) photoionized component and a $kT = 0.9$~keV thermal plasma. These components are likely associated with excitation by the central AGN and shock-heated gas in the surrounding cocoon- and H~II-like regions \citep{trindade_falcao_mapping_2025}. The intrinsic 2–10~keV luminosity is low, $L_{\rm int} \sim 3 \times 10^{39}$~erg~s$^{-1}$, with minimal obscuration ($N_{\rm H} \sim 4 \times 10^{21}$~cm$^{-2}$).

The extended region also requires a photoionized component with lower ionization ($\log U = -0.76$) and lower column density ($\log N_{\rm H} = 19.4$~cm$^{-2}$) than in the core, along with a similar $kT = 0.9$~keV thermal component. These findings are consistent with excitation by a low-luminosity AGN and widespread shocks in the NLR.

Our spectral modeling provides strong evidence that the nucleus of NGC~5005 is unobscured, with column densities well below those expected for Compton-thick sources. Furthermore, the agreement between the observed and [O~III]-predicted X-ray luminosities \citep{trindade_falcao_mapping_2025} rules out a recently switched-off AGN.

The bolometric luminosity inferred from the X-ray data also agrees with that derived from [O~III] emission, supporting the interpretation that NGC~5005 hosts an intrinsically low-luminosity AGN. Adopting a black hole mass of $\log(M_{\rm BH}/M_\odot) = 8.27$, we estimate an extremely low Eddington ratio of $\lambda_{\rm Edd} \sim 5 \times 10^{-6}$. At such low accretion rates, standard accretion models predict the disappearance of the broad-line region. However, this is challenged by previous reports of a broad H$\alpha$ line in NGC~5005’s optical spectrum. Alongside similar findings in other systems such as NGC~3147 \citep{bianchi_hst_2019}, our results suggest that thin accretion structures may persist even at very low accretion rates, posing a challenge to canonical models of radiatively inefficient accretion in low-luminosity AGNs.


\bibliographystyle{aasjournal}
\bibliography{references_zotero}

\end{document}